\documentclass{llncs}  

\usepackage{latexsym}
          
\usepackage{graphicx}
\usepackage{amsmath}
\usepackage{graphics}
\usepackage{amssymb}
\usepackage{algorithmic}
\usepackage{algorithm}
\usepackage{boxedminipage}
\usepackage{booktabs}
\usepackage{url}

\usepackage{epsf}
\usepackage{pgf}
\usepackage{tikz}
\usetikzlibrary{arrows,automata}

\def\punto{\hspace*{\fill}\Box}
\def\qed{\hspace*{1pt}\Box}

\newcommand{\squishlist}{
   \begin{list}{$\bullet$}
    { \setlength{\itemsep}{0pt}      \setlength{\parsep}{1.75pt}
      \setlength{\topsep}{3pt}       \setlength{\partopsep}{0pt}
      \setlength{\leftmargin}{1.5em} \setlength{\labelwidth}{1em}
      \setlength{\labelsep}{0.5em} } }

\newcommand{\squishend}{
    \end{list}  }

\newcommand{\set}[2]{\ensuremath{\{$ $#1$ $|$ $#2$ $\}}}
\newcommand{\setone}[1]{\ensuremath{\{#1\}}}

\begin{document}
\sloppy

\title{On Chase Termination Beyond Stratification}
\subtitle{[Technical Report and Erratum]}

%\numberofauthors{3} 

\author{Michael Meier\thanks{The work of this author was funded by DFG grant GRK 806/03.}\ \ \ \ \ \ \ \ \ \ \ \ \ \ Michael Schmidt$^*$\ \ \ \  \ \ \ \ \ \ \ \ \ \ Georg Lausen}
\institute{University of Freiburg\\
Institute for Computer Science\\
Georges-K\"{o}hler-Allee, Building 051\\
79110 Freiburg i. Br., Germany\\
\{meierm, mschmidt, lausen\}@informatik.uni-freiburg.de}

\date{\today}

\maketitle

%%% sections
\begin{abstract}
We study the termination problem of the chase algorithm, a central tool
in various database problems such as the constraint implication problem,
Conjunctive Query optimization, rewriting queries using views, data exchange, 
and data integration. The basic idea of the chase is, given a database
instance and a set of constraints as input, to fix constraint violations
in the database instance. It is well-known that, for an arbitrary set
of constraints, the chase does not necessarily terminate (in general,
it is even undecidable if it does or not). Addressing this issue, we
review the limitations of existing sufficient termination conditions for
the chase and develop new techniques that allow us to establish weaker
sufficient conditions. In particular, we introduce two novel
termination conditions called {\it safety} and {\it inductive restriction},
and use them to define the so-called {\it $T$-hierarchy} of termination
conditions. We then study the interrelations of our termination conditions
with previous conditions and the complexity of checking our conditions.
This analysis leads to an algorithm that checks membership in a
level of the $T$-hierarchy and accounts for the complexity of termination
conditions. As another contribution, we study the problem of {\it data-dependent}
chase termination and present sufficient termination conditions w.r.t.~fixed
instances. They might guarantee termination although the chase does
not terminate in the general case.
As an application of our techniques beyond those already mentioned,
we transfer our results
into the field of query answering over knowledge bases where the chase
on the underlying database may not terminate, making existing 
algorithms applicable to broader classes of constraints.
\end{abstract}

\section{Introduction}

The chase procedure is a fundamental algorithm that has been successfully
applied in a variety of database
applications~\cite{mms1979,jk1982,bv1984,h2001,l2002,fkmp2005,fkmt2006,dpt2006,ohk2009}.
Originally proposed to tackle the implication problem for data
dependencies~\cite{mms1979,bv1984} and to optimize Conjunctive
Queries (CQs) under data dependencies~\cite{asu1979,jk1982}, it has become
a central tool in Semantic Query Optimization (SQO)~\cite{pt1999,dpt2006,sml2008}.
For instance, the chase can be used to enumerate minimal CQs under a set of
dependencies~\cite{dpt2006}, thus supporting the search for more efficient
query evaluation plans. Beyond SQO, the chase algorithm has been applied
in many other contexts, such as data exchange~\cite{fkmp2005},
peer data exchange \cite{fkmt2006}, data integration~\cite{l2002}, query
answering using views~\cite{h2001}, and probabilistic databases~\cite{ohk2009}.

The core idea of the chase algorithm is simple: given a set of dependencies
(also called constraints) over a database schema and a fixed database instance
as input, it fixes constraint violations in the instance. As a minimal and
intuitive scenario we consider a database graph schema that provides a relation
$E(\textit{src},\textit{dst})$, which stores directed edges from node \textit{src}
to node \textit{dst}, and a node relation $S(n)$ containing nodes with
some distinguished properties, which are enforced by constraints. These
constraints will vary from example to example and we denote nodes in $S$ as
{\it special nodes} in the following. We sketch the idea of the chase algorithm
using a single constraint $\alpha_1:=\forall x( S(x) \rightarrow \exists y E(x,y))$,
stating that each special node has at least one outgoing edge. Now consider
the sample database instance $I := \{ S(n_1), S(n_2), E(n_1,n_2) \}$.
It is easy to see that~$I$ does not satisfy $\alpha_1$, because
it does not contain an outgoing edge for special node $n_2$.
In its effort to fix the constraint violations in the database instance,
the chase procedure would create the tuple
$t_1 := E(n_2,x_1)$, where $x_1$ is a fresh null value.
The resulting database instance $I' := I \cup \{ t_1 \}$ now satisfies
constraint~$\alpha_1$, so the chase terminates and returns $I'$ as result.

One major problem with the chase algorithm, however, is that it does not
terminate in the general case. To give an idea of the problem, let us
sketch a scenario that induces a non-terminating chase sequence. We
replace the constraint $\alpha_1$ from before by constraint
$\alpha_2:=\forall x (S(x) \rightarrow \exists y E(x,y), S(y))$, which
asserts that each special node links to another special node. Now
consider the instance $I$ from before. Obviously, $I$ does not
satisfy $\alpha_2$, because special node $n_2$ has no outgoing edge.
In response, the chase fixes this constraints violation by adding
the two tuples $E(n_2,x_1)$ and $S(x_1)$ to $I$, where $x_1$ is a
fresh null value. Constraint $\alpha_2$ is then fixed w.r.t.~value
$n_2$, but now the special node $x_1$ introduced in the last chase
step violates $\alpha_2$. In subsequent steps the chase would add
$E(x_1,x_2)$, $S(x_2)$, $E(x_2,x_3)$, $S(x_3)$, $\dots$, where
$x_2$, $x_3$, $\dots$ are fresh null values. Hence, when given
instance $I$ and constraint $\alpha_2$ as input, the chase procedure
will never terminate.

As shown in~\cite{dnr2008}, in general it is undecidable if the chase
terminates or not, even for a fixed instance. Still, addressing
the issue of non-terminating chase sequences, several {\it sufficient}
conditions for the input constraints have been proposed that guarantee
termination on every database instance~\cite{fkmp2005,dnr2008,sml2008,msl2009}.
The common idea is to statically assert that there are no positions in
the database schema where fresh null values might be cyclically created in.
The term {\it position} refers to a position in a relational predicate,
e.g.~$E({\it src},{\it dst})$ has two positions, namely {\it src},
denoted as~$E^1$, and {\it dst}, denoted as~$E^2$. Likewise, we denote
by $S^1$ the only position in predicate $S$. The non-terminating
chase sequence discussed before, for instance, cyclically creates
fresh null values in positions $E^1$ and $S^1$.

One well-known termination condition is {\it weak acyclicity}~\cite{fkmp2005}.
Roughly spoken, it implements a global study of the input
constraints, to detect cyclically connected positions in the constraint set
that introduce some fresh null values. In~\cite{dnr2008}, {\it stratification} 
was introduced which meant to generalize weak acyclicity, claiming that it
suffices to assert weak acyclicity locally for subsets of constraints that
might cyclically cause to fire each other. We will show that stratification, unlike stated by the authors of \cite{dnr2008}, does not generally ensure the termination of the chase, yet, as a central contribution, we can prove that it ensures the termination of at least one chase sequence. Moreover, we show that this sequence can be statically determined from the chase graph. It is important to notice that
the techniques introduced in~\cite{fkmp2005,dnr2008} take only the
constraints into account and not the database instance. We therefore call
such termination conditions {\it data-independent}; their result is either the
guarantee that the chase with these constraints terminates for {\it every}
database instance or that no predictions can be made.

This paper explores sufficient termination conditions beyond the corrected version of stratification,
which (by the best of our knowledge) is the most general termination condition
known so far. As one major contribution, we study data-independent chase termination
and present conditions that generalize stratification. Complementary,
we consider the novel problem of data-dependent chase termination, where our
goal is to derive chase termination guarantees w.r.t.~a fixed instance. In
the remainder of the Introduction we summarize the key concepts and ideas of
our analysis and survey the main results.

{\bf Data-independent chase termination.}
As discussed before, the source of non-terminating chase sequences
are fresh null values that are cyclically created at runtime in some
position(s). We develop new techniques that allow us to statically
approximate the set of positions where null values are created in or
copied to during chase application and use them to develop a hierarchy
of sufficient termination conditions that are strictly more general
than stratification. Our termination conditions rely on the following ideas.

{\it (1)~Correction and exploration of the stratification condition:}
We show that stratification does not generally ensure termination of every chase sequence, as stated by the authors of \cite{dnr2008}, but of at least one chase sequence. Besides, we show that such a sequence can be statically determined independently of the input instance. This opens the door to the area of sufficient termination conditions for the chase that ensure, independently of the underlying data, the termination of at least one chase sequence and not necessarily of all. Furthermore, we propose a possible correction of the stratification condition which ensures the termination for every chase sequence, as intended by the authors of \cite{dnr2008}, using the oblivious chase.

{\it (2)~Identification of harmless null values:} Often constraints 
introduce fresh null values in a certain position, but the (fixed) size of the
database instance implies an upper bound on the number of null values that might
be introduced in this position. Consider for example the constraint
$\alpha_3 := \forall x, y (S(x), E(x,y) \rightarrow \exists z E(z,x))$, which
may create fresh null values in position $E^1$. Whenever $\alpha_3$ is part of a
constraint set that does not copy null values to or create null values in
position $S^1$, the number of fresh null values that might be introduced
in position $E^1$ by $\alpha_3$ is implicitly fixed by the number of
entries in relation $S$ and constraint $\alpha_3$ cannot cause an infinite
cascading of fresh null values in this position.

{\it (3)~Supervision of the flow of null values:}
We statically approximate the set of positions where null values might be
copied to during chase application, by a sophisticated study of the interrelations
between the individual constraints. Again, we illustrate the idea by a small and
simple example. Let us consider the two constraints
$\beta_1 := \forall x, y (S(x), E(x,y) \rightarrow E(y,x))$
and $\beta_2 := \forall x, y (S(x), E(x,y) \rightarrow \exists z E(y,z), E(z,x))$,
which assert that each special node with an outgoing edge has cycles of length
$2$ and $3$, respectively. We observe that none of these constraints inserts fresh null values into
relation $S$, so the chase will terminate as soon as
$\beta_1$ and $\beta_2$ have been fixed for all special nodes with an outgoing
edge, i.e.~after a finite number of steps.
Somewhat surprisingly, none of the existing conditions recognizes chase termination
for the above scenario. The reason is that they do not supervise the flow
of null values. Our approach exhibits such an analysis and would guarantee
chase termination for the two constraints above.

{\it (4)~Inductive decomposition of the constraint set:}
The constraint set in the previous example is not dangerous, because
no fresh null values are created in position $S^1$. Let us, in addition
to $\beta_1$ and $\beta_2$, consider the constraint
$\beta_3 := \exists x, y S(x), E(x,y)$, 
stating that there is at least one special node with an outgoing edge.
Clearly, $\beta_3$ fires at most once, so the chase for the constraint
set $\{ \beta_1, \beta_2, \beta_3 \}$ will still terminate. However,
$\beta_3$ complicates the analysis because it ``infects'' position
$S^1$ in the sense that now null values may be created in this position.
We resolve such situations by an (inductive) decomposition of the
constraint set. When applied to the above example, our approach would recognize
that $\beta_3$ is not cyclically connected with $\beta_1$ and $\beta_2$,
and decompose the constraint set into the subsets $\{ \beta_1, \beta_2 \}$
and $\{ \beta_3 \}$, which then are inspected recursively.

%%%%%%%%%%%%%%%%%%%%%%%%%%%%%%%%%%%%%%%%%%%%%%%%%%%%%%
\begin{figure}[t]
\begin{center}
\includegraphics[height=8cm]{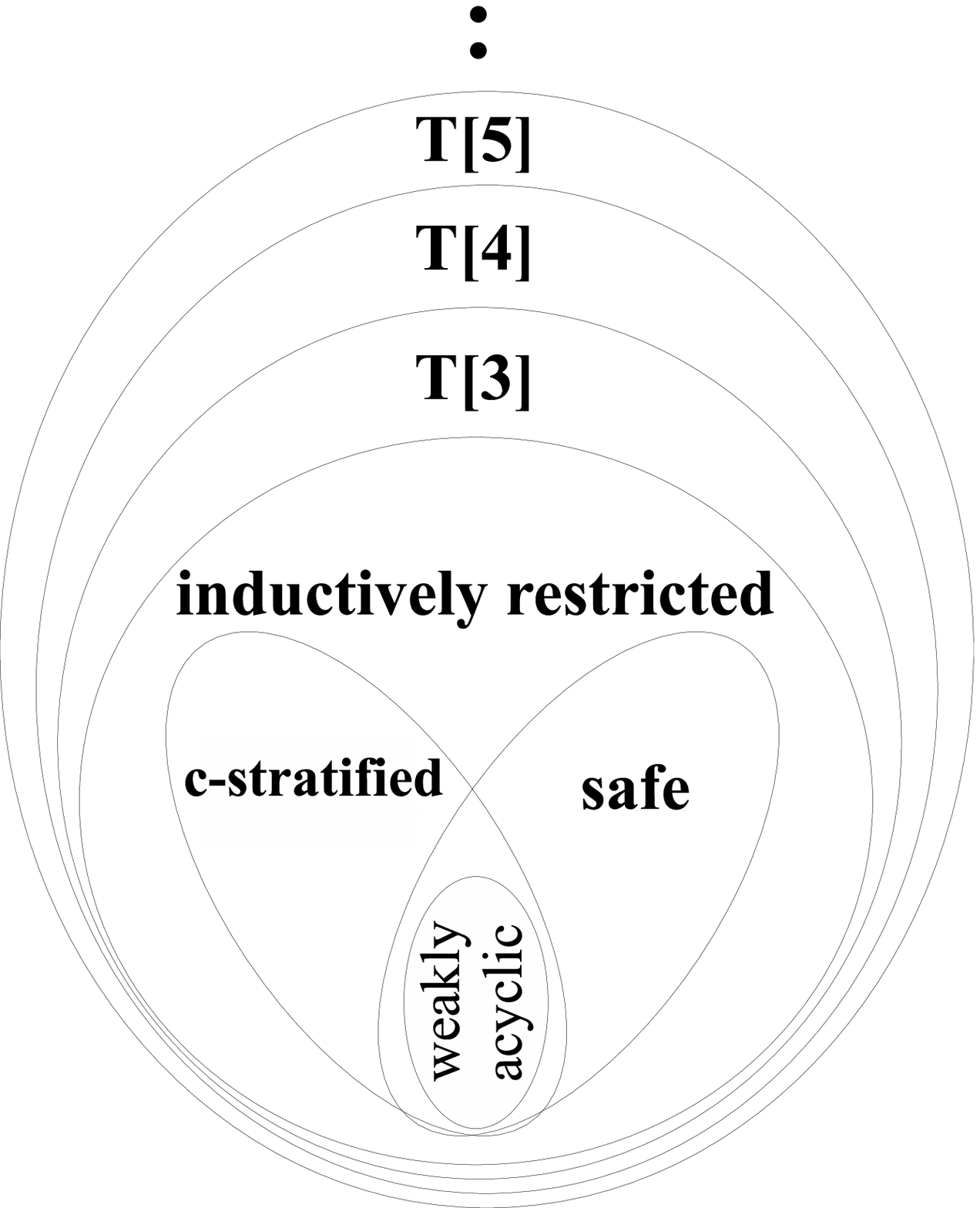}
\end{center}
%\vspace{-0.3cm}
\caption{Chase termination conditions.}
%\vspace{-0.4cm}
\label{fig:survey}
\end{figure}

%%%%%%%%%%%%%%%%%%%%%%%%%%%%%%%%%%%%%%%%%%%%%%%%%%%%%%

Based upon the previous ideas we develop two novel sufficient chase
termination condition, called {\it safety} and {\it inductive restriction}.
Figure~\ref{fig:survey} surveys our main results and relates them to the
previous termination condition weak acyclicity and the corrected version of 
stratification that we call  c-stratification. All
classes in the figure guarantee chase termination in polynomial-time
data complexity and all inclusion relationships are strict. As can be seen,
safety generalizes weak acyclicity and is further generalized by inductive
restriction. On top of inductively restricted constraints we then
define an (infinite) hierarchy of sufficient termination conditions, which we
call $T$-hierarchy. To give an intuition, for a fixed level in this hierarchy,
say $T[k]$, the idea is to study the flow and creation of fresh null values detailedly
for chains of up to~$k$ constraints that might cause to fire each other in sequence.

{\bf An algorithm.}
It can be checked in polynomial time if a constraint set is safe; in contrast,
the recognition problem for inductively restricted constraints and the classes in the
$T$-hierarchy is in~ \textsc{coNP}. We develop an efficient algorithm that
accounts for the increasing complexity of the recognition problem and can be
used to test membership of a constraint set in some fixed level of the $T$-hierarchy.
The underlying idea idea of our algorithm is to combine the
different sufficient termination conditions, to reduce the
complexity of checking for termination wherever possible.

{\bf Data-dependent chase termination.}
Whenever the input constraint set does not fall into some fixed level of the
$T$-hierarchy, no termination guarantees for the general case can be derived.
Arguably, reasonable applications should never risk non-termination, so the chase
cannot be safely applied to {\it any instance} in this case. Tackling this
situation, we study the novel problem of {\it data-dependent chase termination:}
given constraint set $\Sigma$ and a fixed instance~$I$, does the chase with
$\Sigma$ terminate on~$I$? We argue that this setting particularly makes sense
in the context of Semantic Query Optimization, where the query -- interpreted
as database instance -- is chased: typically, the query is small, so the
``data'' part can be analyzed efficiently (as opposed to the case where the
input is a large database instance).  We propose two complementary approaches:

\begin{enumerate}
\item Our first, static scheme relies on the observation that, if the instance is fixed, we
can ignore constraints in the constraint set that will never fire when chasing
the instance, i.e.~if general sufficient termination guarantees hold
for those constraints that might fire. As a fundamental result, we show that in general
it is undecidable if a constraint will never fire on a fixed instance.
Still, we give a {\it sufficient} condition that allows us to identify
such constraints in many cases and derive a sufficient data-dependent 
condition.

\item Whenever the static approach fails, our second, dynamic approach comes into play:
we run the chase and track cyclically created fresh null values 
in a so-called monitor graph. We then fix the maximum depth of cycles in the
monitor graph and stop the chase when this limit is exceeded: in such a case,
no termination guarantees can be made. However, we show that each fixed search
depth implicitly defines a class of constraint-instance pairs for which the
chase terminates. Intuitively, the search depth limit can be seen as a natural
condition that allows us to stop the chase when ``dangerous'' situations arise.
Under these considerations, our approach adheres to situations that are likely
to cause non-termination, so it is preferable to blindly running the chase
and aborting after a fixed amount of time, or a fixed number of chase steps.
Applications might fix the search depth following a pay-as-you-go principle.
Ultimately, the combination of our static and dynamic analysis constitutes
a pragmatic workaround in all scenarios where no general (i.e., data-independent)
termination guarantees can be made. 
\end{enumerate}

{\bf Application.}
As a possible application of our techniques, we review the problem of answering
Conjunctive Queries over knowledge bases in the presence of constraints, with
a focus on scenarios where the chase with the given constraint set
does not necessarily terminate. This problem was first considered in~\cite{jk1982}
and recently generalized in~\cite{cgk2008,cgl2009}. A key idea in~\cite{cgk2008}
is an overestimation of the set of positions in which null values might occur,
using the concept of so-called affected positions. In particular, affected
positions are used in~\cite{cgk2008} to define a class of constraints
called weakly guarded constraint sets, for which the query answering problem
is decidable. Using our novel techniques,
we refine the notion of affected positions with the help of a so-called
restriction system, which is a central tool in our study of data-independent
chase termination, e.g.~used to define the class of inductively restricted  
constraints and the $T$-hierarchy. We show that restriction systems can be fruitfully
applied to generalize the class of weakly guarded constraints to a class we
call restrictedly guarded constraints, thus making the algorithms
in \cite{cgk2008,cgl2009} applicable to a larger class of constraints.

{\bf Structure.} 
Section \ref{sec:prel} presents the necessary background in databases. Next,
Section~\ref{sec:ind} provides our results on data-independent chase termination.
Its main results are the exploration/correction of the stratification condition, the introduction of the $T$-hierarchy and an algorithm to
efficiently test membership of a constraint set in some level of the $T$-hierarchy.
In Section~\ref{sec:dep} we then motivate the novel problem of
data-dependent chase termination. As a possible application, Section~\ref{sec:apps} demonstrates
the applicability of our concepts and methods in the context of query answering
on knowledge bases where the chase may not terminate. We conclude with
some closing remarks in Section~\ref{sec:conclusion}. 

{\bf Additional remarks.} This paper builds upon the ideas presented in
the Extended Abstract~\cite{msl2_2009}. Other parts of this paper were
informally published as technical reports~\cite{sml2008,msl2009}. 
%None of~\cite{sml2008,msl2009,msl2_2009}
%violates the duplicate submission policy of VLDB.

\section{Preliminaries}
\label{sec:prel}

{\bf General mathematical notation.} The natural numbers $\mathbb{N}$ do not include $0$. 
For $n \in \mathbb{N}$, we denote by $[n]$ the set $\setone{1,...,n}$. For a set $M$,
we denote by $2^M$ its powerset and by $|M|$ its cardinality. Abusing notation we denote
by $|s|$ also the length of a logical formula. Given a tuple $t = (t_1,\dots,t_n)$ we define the
tuple obtained by projecting on positions $1 \leq i_1 < \dots < i_m \leq n$ as
$p_{i_1,\dots,i_m}(t) := (t_{i_1},\dots,t_{i_m})$. \\

\textbf{Databases.} We fix three pairwise disjoint infinite sets:
the set of \textit{constants}~$\Delta$, the set of \textit{labeled nulls}
$\Delta_{null}$, and the set of \textit{variables}~$V$. Often we will denote a
sequence of variables, constants or labeled nulls by $\overline{a}$ if the
length of this sequence is understood from the context. A \textit{database schema}
$\mathcal{R}$ is a finite set of relational symbols $\setone{R_1,...,R_n}$. To every
relational symbol $R \in \mathcal{R}$ we assign a natural number $ar(R)$
called its {\it arity}. A database position is a pair $(R,i)$ where
$R \in \mathcal{R}$ and $i \in [ar(R)]$, for short we write $R^i$, e.g.~a three-ary
predicate $S$ has three positions $S^1, S^2, S^3$. We say that a variable, labeled
null, or constant $c$ appears e.g.~in position $R^1$ if there exists an atom
$R(c,...)$. In the rest of the paper, we assume the database schema and the set of
constants and labeled nulls to be fixed and therefore we will suppress them in our
notations.  A \textit{database instance} $I$ is a finite set of $\mathcal{R}$-atoms
that contains only elements from  $\Delta \cup \Delta_{null}$ in its positions.
The domain of $I$, $dom(I)$, is the set of elements from $\Delta \cup \Delta_{null}$
that appear in $I$. \\

 \textbf{Conjunctive Queries.} 
 A Conjunctive Query (CQ) is an expression of the form
$ans(\overline{x}) \leftarrow \varphi(\overline{x},\overline{z})$,
where $\varphi$ is a conjunction of relational atoms, $\overline{x}$,
$\overline{z}$ are sequences of variables and constants, and it
holds that every variable in $\overline{x}$ also occurs in $\varphi$.
If $\overline{x}$ is empty we call the query boolean.
The semantics of such a query $q$ on database instance $I$ is defined as
$q(I) := \set{\overline{a} \in \Delta^{|\overline{x}|}}{I \models \exists \overline{z} \varphi(\overline{a},\overline{z})}$.\\

%is an $n$-tuple $(I_1,...,I_n)$,
%where $I_i \subseteq (\Delta \cup \Delta_{null})^{ar(R_i)}$ for every $i \in [n]$.
%We write {\it dom}($I$) for the set of values in $I$ and
%will denote $(c_1,...,c_{ar(R_i)}) \in I_i$ by the \textit{fact} $R_i(c_1,...,c_{ar(R_i)})$
%and shall represent the instance $I$ as the set of its facts.
%Abusing notation, we write $I = \set{R_i(t)}{t \in I_i, i \in [n]}$.

\textbf{Constraints.} Let $\overline{x}, \overline{y}$ be sequences of variables.
We consider two types of database constraints: \textit{tuple generating dependencies} 
(TGDs) and \textit{equality generating dependencies} (EGDs). A TGD is a first-order
sentence
$\alpha := \forall \overline{x} (\phi(\overline{x}) \rightarrow \exists \overline{y} \psi(\overline{x},\overline{y}))$ such that~(a) both $\phi$ and $\psi$ are conjunctions of atomic formulas (possibly with parameters from $\Delta$), (b) $\psi$ is not empty, (c) $\phi$ is possibly empty, (d) both $\phi$ and
$\psi$ do not contain equality atoms and (e) all variables from $\overline{x}$ that occur
in $\psi$ must also occur in $\phi$. We denote by $pos(\alpha)$ the set of
positions in $\phi$. An EGD is a first-order sentence
$\alpha := \forall \overline{x} (\phi(\overline{x}) \rightarrow x_i = x_j)$,
 where $x_i, x_j$ occur in $\phi$ and $\phi$ is a non-empty conjunction of equality-free
$\mathcal{R}$-atoms (possibly with parameters from $\Delta$). 
We denote the set of positions in $\phi$ by $pos(\alpha)$.

From now on we will use the word constraint instead of saying that a logical expression may be a TGD or an EGD. Satisfaction of constraints by databases is defined in the standard first-order manner and is therefore omitted here. We write $I \models \alpha$ if a constraint $\alpha$ is satisfied by $I$ and $I \not \models \alpha$ otherwise.
As a notational convenience, we will often omit the $\forall$-quantifier and
respective list of universally quantified variables. For a set of TGDs and EGDs
$\Sigma$ we set $pos(\Sigma) := \bigcup_{\xi \in \Sigma} pos(\xi)$. We use the term $body(\alpha)$ for a constraint $\alpha$ as the set of atoms in its premise; analogously we define $head(\alpha)$.
In case $\alpha$ is a constraint and $\overline{a}$ is a sequence of labeled nulls and constants, then $\alpha(\overline{a})$ is the constraint $\alpha$ without universal quantifiers but with parameters $\overline{a}$. We will often abuse this notation and say that a labeled null occurs in $\alpha(\overline{a})$, meaning that a labeled null
is the parameter for some universally quantified variable in $\alpha$.\\

%\textbf{Homomorphisms.}
%A homomorphism from a set of atoms $A_1$ to a set of atoms $A_2$ is a mapping
%$\mu : \Delta \cup \Delta_{null} \cup V \rightarrow \Delta \cup \Delta_{null} \cup V$
%s.t.~the following conditions hold: (a) if $c \in \Delta$, then $\mu(c)=c$, (b)
%if $c \in \Delta_{null}$, then $\mu(c) \in \Delta \cup \Delta_{null}$, and (c)~if
%$R(c_1,...,c_n) \in A_1$, then $R(\mu(c_1),...,\mu(c_n)) \in A_2$.

\textbf{Homomorphisms.}
A homomorphism from a set of atoms $A_1$ to a set of atoms $A_2$ is a mapping
$\mu : \Delta \cup V \rightarrow \Delta \cup \Delta_{null}$
such that the following conditions hold: (i) if $c \in \Delta$, then $\mu(c)=c$ and (ii)~if $R(c_1,...,c_n) \in A_1$, then $R(\mu(c_1),...,\mu(c_n)) \in A_2$.\\

\textbf{Chase.}
Let $\Sigma$ be a set of TGDs and EGDs and $I$ an instance, represented as a set of atoms. We say that a TGD $\forall \overline{x} \varphi \in \Sigma$ is applicable to $I$ if there is a homomorphism $\mu$ from $body(\forall \overline{x} \varphi)$ to~$I$ and $\mu$ cannot be extended to a homomorphism $\mu' \supseteq \mu$ from $head(\forall \overline{x}\varphi)$ to $I$. In such a case the chase step $I \stackrel{\forall \overline{x}\varphi, \mu(\overline{x})}{\longrightarrow} J$ is defined as follows. We define a homomorphism $\nu$ as follows: (a)~$\nu$ agrees with $\mu$ on all universally quantified variables in $\varphi$, (b) for every existentially quantified variable $y$ in $\forall \overline{x} \varphi$ we choose a "fresh" labeled null $n_y \in \Delta_{null}$ and define $\nu(y):=n_y$. We set $J$ to be $I \cup \nu(head(\forall \overline{x}\varphi))$. We say that an EGD $\forall \overline{x}\varphi \in \Sigma$ is applicable to~$I$ if there is a homomorphism $\mu$ from $body(\forall \overline{x}\varphi)$ to $I$ and it holds that $\mu(x_i) \neq \mu(x_j)$. In such a case the chase step $I \stackrel{\forall \overline{x}\varphi, \mu(\overline{x})}{\longrightarrow} J$ is defined as follows. We set $J$ to be 
\squishlist
	\item $I$ except that all occurrences of $\mu(x_j)$ are substituted by $\mu(x_i) =: a$, if $\mu(x_j)$ is a labeled null,
	\item $I$ except that all occurrences of $\mu(x_i)$ are substituted by $\mu(x_j) =: a$, if $\mu(x_i)$ is a labeled null,
	\item undefined, if both $\mu(x_j)$ and $\mu(x_i)$ are constants. In this case we say that the chase fails.
\squishend

A chase sequence is an exhaustive application of applicable constraints 
$I_0 \stackrel{\varphi_0, \overline{a}_0}{\longrightarrow} I_1 \stackrel{\varphi_1, \overline{a}_1}{\longrightarrow} \ldots$,
 where we impose no strict order what constraint must be applied in case several constraints apply. If this sequence is finite, say $I_r$ being its final element, the chase terminates and its result $I_0^{\Sigma}$ is defined as $I_r$. The length of this chase sequence is $r$. Note that different orders of application of applicable constraints may lead to a different chase result. However, as proven in \cite{fkmp2005}, two different chase orders lead to homomorphically equivalent results, if these exist. Therefore, we write $I^{\Sigma}$ for the result of the chase on an instance $I$ under constraints $\Sigma$. It has been shown in \cite{mms1979,bv1984,jk1982} that $I^{\Sigma} \models \Sigma$. In case that a chase step cannot be performed (e.g., because a homomorphism would have to equate two constants) the chase result is undefined. 
 If we have an infinite chase sequence $I_0 \stackrel{\varphi_0, \overline{a}_0}{\longrightarrow} I_1 \stackrel{\varphi_1, \overline{a}_1}{\longrightarrow} \ldots$, we distinguish two cases: (i) if the constraint set contains an EGD, then we also say that the result is undefined; (ii) if the constraint set consists of TGDs only then $I^{\Sigma} := \bigcup_{i \geq 0} I_i$ is the union of all intermediate database instances during the application of the chase.\\
 
 \textbf{Oblivious Chase.} We will also use oblivious chase steps throughout this paper. An oblivous chase step for a TGD $\forall \overline{x} \varphi$ is defined as follows. The oblivious step applies to an instance $I$ if there is a homomorphism $\mu$ from $body(\forall \overline{x} \varphi)$ to~$I$. In such a case the oblivious chase step $I \stackrel{*,\forall \overline{x}\varphi, \mu(\overline{x})}{\longrightarrow} J$ is defined as follows. We define a homomorphism $\nu$ as follows: (a)~$\nu$ agrees with $\mu$ on all universally quantified variables in $\varphi$, (b) for every existentially quantified variable $y$ in $\forall \overline{x} \varphi$ we choose a "fresh" labeled null $n_y \in \Delta_{null}$ and define $\nu(y):=n_y$. We set $J$ to be $I \cup \nu(head(\forall \overline{x}\varphi))$. An oblivious chase step for an EGD is a chase step for an EGD except that we also add an $*$ on the arrow (like in the case of TGDs) that indicates the step. Intuitively, an oblivious chase step always applies when the body of a constraint can be mapped to an instance, even if the constraint is satisfied.\\

\section{Data-independent Termination}
\label{sec:ind}

In this section we discuss the sufficient data-independent chase termination
conditions presented in Figure~\ref{fig:survey}.
First, we will review existing approaches and then introduce the novel
class of {\it safe} constraints, which strictly generalizes weak acyclicity,
but is different from stratification. Building upon the definition of safety,
we then introduce {\it inductively restricted} constraints as a consequent
advancement of our ideas. The latter class strictly subsumes all termination
conditions known so far. Finally, we will define a hierarchy of sufficient
termination condition on top of inductively restricted constraints,
the so-called $T$-hierarchy.
Each level $T[k]$ in this hierarchy is strictly contained in the next level
$T[k+1]$. Our novel sufficient termination conditions vastly extend the
applicability of the chase algorithm, as they guarantee chase
termination for much larger classes of constraints than previous conditions.

\begin{figure}[t]
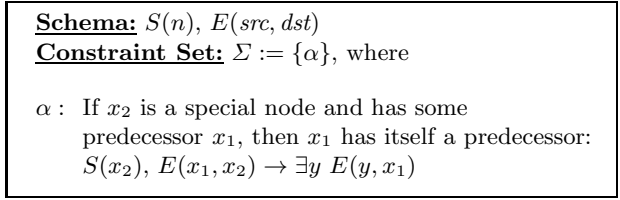

\centering
\begin{boxedminipage}{6.5cm}
\begin{tabbing}
x \= \kill
\>\underline{\bf Schema:} $S(n)$, $E(\textit{src},\textit{dst})$\\
\>\underline{\bf Constraint Set:} $\Sigma := \setone{\alpha}$, where
\end{tabbing}
%\vspace{-0.3cm}
\begin{tabbing}
x \= xxx \= \kill
\>$\alpha:$\>If $x_2$ is a special node and has some\\ 
\>\>predecessor $x_1$, then $x_1$ has itself a predecessor:\ \ \ \\
\>\>$S(x_2)$, $E(x_1,x_2) \rightarrow \exists y\ E(y,x_1)$\\[-0.3cm]
\end{tabbing}
\end{boxedminipage}
%\vspace{-0.3cm}
\caption{A sample constraint.}
%\vspace{-0.4cm}
\label{fig:schemaandconstraints1}
\end{figure}

As a minimalistic motivating example for our study of novel chase termination
conditions let us consider the constraint set $\Sigma$ from
Figure~\ref{fig:schemaandconstraints1}, which is settled in our graph
database schema from the Introduction. As we shall see later, the chase with
$\Sigma$ terminates for every database instance. Still, none of the existing
termination conditions is able to recognize termination for this constraint set,
i.e.~$\Sigma$ is neither
weakly acyclic nor stratified. With the techniques and tools that we develop
within this section, we will be able to guarantee chase termination for
$\Sigma$ on every database instance.

\subsection{Weak Acyclicity}
The notion of weak acyclicity from~\cite{dt2001,fkmp2005} is the starting point for
our discussion. Informally spoken, the key idea of weak acyclicity is to
statically estimate the flow of data between the database positions
during the execution of the chase. Weak acyclicity asserts that no
fresh values are created over and over again.

\begin{definition} \em (see \cite{fkmp2005})
The dependency graph $\mbox{dep}(\Sigma)$ of a set of constraints $\Sigma$ is the directed graph defined as follows. The set of vertices is the set of positions that occur in some TGD in $\Sigma$. There are two kinds of edges. Add them as follows: for every TGD 
$\forall \overline{x} (\phi(\overline{x}) \rightarrow \exists \overline{y} \psi(\overline{x},\overline{y})) \in \Sigma$
 and for every $x$ in $\overline{x}$ that occurs in $\psi$ and every occurrence of $x$ in $\phi$ in position $\pi_1$
  
\squishlist
	\item for every occurrence of $x$ in $\psi$ in position $\pi_2$, add an edge $\pi_1 \rightarrow \pi_2$ (if it does not already exist).
	\item for every existentially quantified variable $y$ and for every occurrence of $y$ in a position $\pi_2$, add a {\it special} edge $\pi_1 \stackrel{*}{\rightarrow} \pi_2$ (if it does not already exist).
\squishend
A set $\Sigma$ of TGDs and EGDs is called \textit{weakly acyclic} iff $\mbox{dep}(\Sigma)$ has no cycles going through a special edge. $\punto$
\label{def:wa}
\end{definition}

Intuitively, normal edges in the dependency graph track the flow of data
between the database positions and special edges cover the case of
newly introduced null values.
If the dependency graph contains no cycles through a special edge it
cannot happen that fresh null values are cyclically added to the database
instance. It has been shown in~\cite{fkmp2005} that weak acyclicity can be
decided in polynomial time. We illustrate the definition of weak acyclicity
by example.
%for a situation in which weak acyclicity can guarantee the termination of the chase.

\begin{figure}[t]
\begin{center}
\includegraphics[height=3cm]{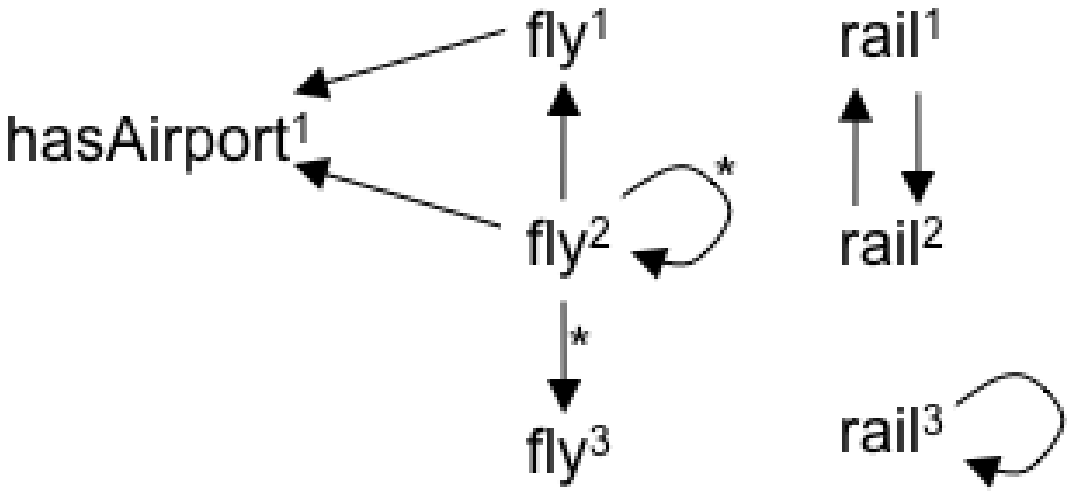}
\end{center}
%\vspace{-0.5cm}
\caption{Dependency graph for $\Sigma$ from Figure~\ref{fig:schemaandconstraints}.}
%\vspace{-0.4cm}
\label{fig:weak1}
\end{figure}

\begin{example} \label{ex:weak1} \em 
We depict the dependency graph for the constraint set
$\Sigma := \{ \alpha_1, \alpha_2, \alpha_3\}$
from Figure~\ref{fig:schemaandconstraints} in Figure~\ref{fig:weak1}.
One can observe that $\Sigma$ is not weakly acyclic, as witnessed by the self-loop through
special edge $\textit{fly}^2 \stackrel{*}{\rightarrow} \textit{fly}^2$.$\punto$
\end{example}

\subsection{Stratification}
In \cite{dnr2008}, stratification was introduced which meant to improve the former weak acyclicity condition.
The main idea behind stratification is to decompose the constraint set into
independent subsets that are then separately tested for weak acyclicity.
More precisely, the decomposition splits the input constraint set into
subsets of constraints that may cyclically cause to
fire each other. The idea is that the termination guarantee for the full constraint set
should follow if weak acyclicity holds for each subset in the decomposition.

\begin{definition} \em (see \cite{dnr2008})
Given two TGDs or EGDs $\alpha, \beta \in \Sigma$ we define $\alpha \prec \beta$ iff there exists a relational database instance $I$ and $\overline{a}, \overline{b}$ such that
(i) $I \not \models \alpha(\overline{a})$, (ii) $I \models \beta(\overline{b})$, (iii) $I \stackrel{\alpha, \overline{a}}{\rightarrow} J$, and (iv)~$J \not \models \beta(\overline{b})$. $\punto$
\end{definition}

Intuitively, $\alpha \prec \beta$ means that if $\alpha$ fires it can cause $\beta$ to fire (in the case that $\beta$ could not fire before). We give an example to illustrate this definition.

\begin{example} \em (see \cite{dnr2008}) Let predicate $E$ store the edge relation of a graph
and let the constraint $\alpha := E(x_1,x_2), E(x_2,x_1) \rightarrow \exists y_1, y_2 E(x_1,y_1), E(y_1,y_2), E(y_2,x_1)$ be given, stating that each node having a cycle of length $2$ also has a cycle of length $3$. A $3$-cycle can never be a $2$-cycle again, so it holds that $\alpha \not \prec \alpha$. $\punto$ \label{ex:strat}
\end{example}

The actual definition of stratification then relies, as outlined before, on the notion of weak acyclicity.

\begin{definition} \em(see \cite{dnr2008})
The chase graph $G(\Sigma)=(\Sigma,E)$ of a set of constraints $\Sigma$ contains a directed edge $(\alpha,\beta)$ between two constraints iff $\alpha \prec \beta$. We call $\Sigma$ stratified iff the constraints in every cycle of $G(\Sigma)$ are weakly acyclic. $\punto$
\end{definition}

Stratification strictly generalizes
weak acyclicity (see~\cite{dnr2008}), thus~(i)~if $\Sigma$ is weakly acyclic, then it is also stratified
and (ii)~there are constraint sets that are stratified but not weakly acyclic
(cf.~Example~\ref{ex:stratwa}).

\begin{example}\em
Consider the constraint $\alpha$ from Example~\ref{ex:strat}. It holds that
$\alpha \not \prec \alpha$, so $\setone{\alpha}$ is stratified. As shown in~\cite{dnr2008},
the dependency graph of $\setone{\alpha}$ contains a cycle through a special edge, so
$\setone{\alpha}$ is not weakly acyclic. $\punto$
\label{ex:stratwa}
\end{example}

It can be decided in  $\mbox{coNP}$ whether a set of constraints is stratified. The authors of \cite{dnr2008} claimed the following result:

\begin{claim} \cite{dnr2008} \em
Let $\Sigma$ be a fixed set of stratified constraints. Then, there exists a
polynomial $Q \in \mathbb{N}[X]$ such that for any database instance $I$, the length
of every chase sequence is bounded by $Q(|dom(I)|)$. $\punto$
\end{claim}

Unfortunately, we could show that this claim is wrong as the next example shows.

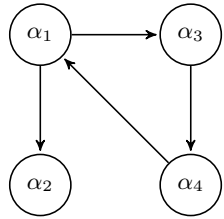
\begin{figure}
\begin{center}
\begin{tabular}{c}

\begin{tikzpicture}[->,>=stealth',shorten >=1pt,auto,node distance=2cm,semithick]
\node[state] (a) [] {$\alpha_1$};
\node[state] (b) [below of=a] {$\alpha_2$};
\node[state] (c) [right of=a] {$\alpha_3$};
\node[state] (d) [below of=c] {$\alpha_4$};
\path (a) edge []          node {$$} (b)
      (a) edge [] node {$$} (c)
      (c) edge [] node {$$} (d)
      (d) edge [] node {$$} (a);
\end{tikzpicture}	
\end{tabular}

\end{center}
\caption{Chase graph for example \ref{ex:stratfalsch}}
\label{fig:falsch}
\end{figure}

\begin{example}\em
Given the set of TGDs $\Sigma = \setone{\alpha_1,...,\alpha_4}$, where
\begin{tabbing}
x \= xxx \= xxx \= \kill
\>$\alpha_1$ \> $:=$ \> $R(x_1) \rightarrow S(x_1,x_1)$,\\
\>$\alpha_2$ \> $:=$ \> $S(x_1,x_2) \rightarrow \exists z T(x_2,z)$,\\
\>$\alpha_3$ \> $:=$ \> $S(x_1,x_2) \rightarrow T(x_1,x_2), T(x_2,x_1)$ and\\
\>$\alpha_4$ \> $:=$ \> $T(x_1,x_2), T(x_1,x_3), T(x_3,x_1) \rightarrow R(x_2)$.
\end{tabbing}

We will give now an instance for which the chase does not necessarily 
terminate. Consider the database $\setone{R(a)}$ and the chase sequence which applies the constraints in the order 
$\alpha_1,...,\alpha_4,\alpha_1,...,\alpha_4,...$ and so on. The first steps of the resulting chase sequence look as follows:

\begin{tabbing}
x \= xxxxxxxxx \= xxxxxxxxxx \= \kill
\>\> $\setone{R(a)}$\\
\>$\stackrel{\alpha_1,a}{\longrightarrow}$ \> $ \setone{R(a), S(a,a)}$\\
\>$\stackrel{\alpha_2,a,a}{\longrightarrow}$ \> $ \setone{R(a), S(a,a), T(a,n_1)}$\\
\>$\stackrel{\alpha_3,a,a}{\longrightarrow} $ \> $\setone{R(a), S(a,a), T(a,n_1),T(a,a)}$\\
\>$\stackrel{\alpha_4,a,n_1,a}{\longrightarrow} $ \> $\setone{R(a), S(a,a), T(a,n_1),T(a,a),R(n_1)}$\\
\>$\stackrel{\alpha_1,n_1}{\longrightarrow} $ \> $\setone{R(a), S(a,a), T(a,n_1),T(a,a),R(n_1),S(n_1,n_1)}$\\
\>$\stackrel{\alpha_2,n_1,n_1}{\longrightarrow} $ \> $\setone{R(a), S(a,a), T(a,n_1),T(a,a),R(n_1),S(n_1,n_1),T(n_1,n_2)}$\\
\>$\stackrel{\alpha_3,n_1,n_1}{\longrightarrow} $ \> $\setone{R(a), S(a,a), T(a,n_1),T(a,a),R(n_1),S(n_1,n_1),T(n_1,n_2),T(n_1,n_1)}$\\
\>$\stackrel{\alpha_4,n_1,n_2,n_1}{\longrightarrow} $ \> $\setone{R(a), S(a,a), T(a,n_1),T(a,a),R(n_1),S(n_1,n_1),T(n_1,n_2),T(n_1,n_1),R(n_2)}$,\\
\>$\stackrel{\alpha_1,a}{\longrightarrow}$ \>\ldots \\
\end{tabbing}

where $n_1,n_2$ are fresh null values. It can be easily seen that this sequence is infinite. The chase graph for $\Sigma$ is depicted in Figure \ref{fig:falsch}. The only cycle in it is constituted by full TGDs only and therefore is weakly acyclic. Hence, it follows that $\Sigma$ is stratified. $\punto$
\label{ex:stratfalsch}
\end{example}

This has profound implications. Unlike weak acyclicity, stratification does not ensure termination of every chase sequence for every instance. However, we can prove another equally useful result with the definition of stratification as in \cite{dnr2008}. If a set of constraints is stratified, we cannot ensure termination for every instance and every chase sequence, but for every instance there is some chase sequence that terminates as stated in the following theorem. We want to emphasize that this result is our own finding.

\begin{theorem} \em
Let $\Sigma$ be a fixed set of stratified constraints. Then, there exists a
polynomial $Q \in \mathbb{N}[X]$ such that for any database instance $I$ there is a terminating chase sequence whose length
is bounded by $Q(|dom(I)|)$. 
\end{theorem}

{\bf Proof:} For the termination see the proof of theorem \ref{thm:stratm}. The polynomial data complexity follows immediately from the polynomial data complexity of weakly acyclic constraint sets. $\punto$

But how can we use this result in practice? The first idea is to apply the chase in a breadth-first manner, i.e.~generating a tree whose root is the start instance, its children are obtained by applying one chase step on the start instance and the tree is expanded in breadth-first manner. This ensures that if there is a terminating chase sequence, then we will find it. Unfortunately, this is rather uneffective  because in some intermediate instance there may be many constraints violated and therefore the degree and the depth of the tree may be high. 

As it turns out, we are in a much better situation here. We can use the chase graph to effectively construct the order in which the constraints must be applied to ensure termination. To the best of our knowledge, this is the first study of sufficient termination conditions for the chase which does not ensure the termination of all chase sequences but at least of some sequence. Additionally, the sequence can be effectively constructed. We give an example to illustrate this.

\begin{example}\em
Consider the constraint set $\Sigma$ from Example~\ref{ex:stratfalsch} again and the instance $\setone{R(a), T(b,b)}$. We give a chase sequence that terminates.

\begin{tabbing}
x \= xxxxxxxxx \= xxxxxxxxxx \= \kill
\>\>$\setone{R(a), T(b,b)}$\\
\>$\stackrel{\alpha_1,a}{\longrightarrow}$ \> $ \setone{R(a), T(b,b), S(a,a)}$\\
\>$\stackrel{\alpha_3,a,a}{\longrightarrow} $ \> $\setone{R(a), T(b,b), S(a,a),T(a,a)}$\\
\>$\stackrel{\alpha_4,b,b,b}{\longrightarrow} $ \> $\setone{R(a), T(b,b), S(a,a),T(a,a), R(b)}$\\
\>$\stackrel{\alpha_1,b}{\longrightarrow} $ \> $\setone{R(a), T(b,b), S(a,a),T(a,a), R(b), S(b,b)}$\\
\end{tabbing}

It holds that $\setone{R(a), T(b,b), S(a,a),T(a,a), R(b), S(b,b)} \models \Sigma$. We obtained a terminating chase sequence by first chasing with the constraints in the cycle and after the chase with these constraints is finished we (possibly) chase with $\alpha_2$, which was not necessary here. It can be shown that this strategy leads always to finite chase sequences, regardless of the underlying instance. Intuitively, this works here because violations of $\alpha_2$ can be repaired with the help of $\alpha_3$. $\punto$
\end{example}

The observations made in this example can be generalized to the next theorem.

\begin{theorem} \em \label{thm:stratm}
Let $\Sigma$ be a fixed set of stratified constraints. If the chase terminates independently of the database instance and independently of the chase order for every strongly-connected component of the chase graph $G(\Sigma)$, then for every database instance $I$ a terminating chase sequence can be effectively constructed. 
\end{theorem}

{\bf Proof:} 
Let the chase graph $G(\Sigma)=(\Sigma,E)$ be given. We write $\alpha \sim \beta$ if and only if $\alpha$ and $\beta$ are contained in a common cycle in $G(\Sigma)$ or $\alpha=\beta$. Note that $\sim$ is an equivalence relation. Let $\Sigma / \sim = \setone{W_1,...,W_n}$ and $E' := \set{(W_i,W_j)}{ i,j \in [n], i \neq j, \text{ there is some } \alpha_i \in W_i, \beta_j \in W_j \text{ such that } \alpha_i \prec \beta_j}$. Let $W'_1,..., W'_n$ be a topological sorting of $(\Sigma / \sim,E')$. Note that $W'_1,..., W'_n$  are the strongly connected components of the chase graph and  constraint sets that are not involved in any cycle in the chase graph, therefore the chase terminates independently of the database instance and independently of the chase order for these constraint sets. Let $I_0$ be an arbitrary database instance. Let $I_i$ be obtained from $I_{i-1}$ by chasing with $W'_i$ for every $i \in [n]$. It holds that $I_2 \models W'_1$. Otherwise there is some $\alpha \in W'_1$, $\beta \in W'_2$ and a database instance $I$ such that $I \models \alpha$, but $I \stackrel{\beta, \overline{b}}{\longrightarrow} J \not \models \alpha$. But this implies $\beta \prec \alpha$ which means $W'_2$ must come before $W'_1$ in the topological sorting of $(\Sigma / \sim,E')$. Using induction on $n$ it can be seen that $I_n \models \Sigma$ (observe that $W'_1,..., W'_n$ is a partition of $\Sigma$). $\punto$

This allows us to apply the chase procedure safely in situations when the termination cannot be guaranteed for every chase sequence. We avoid the overhead of branching in the breadth-first chase and therefore reduce the complexity of generating a chase result. This opens the door to the area of sufficient termination conditions for the chase which ensure, independently of the underlying data, the termination of at least one chase sequence and not necessrily of all. Furthermore, in the next section, we propose a possible correction of the stratification condition which ensures the termination for every chase sequence, using the oblivious chase.

\subsection{C-Stratification}
We propose a possible correction of stratification, called {\it c-stratification}, which has the property of ensuring the termination of the chase independent of the data and the chase sequence used. The underlying ideas are the same as in \cite{dnr2008}: decompose the constraint set such that if a termination guarantee can be made for every subset, then the same guarantee can be made for the overall set.

\begin{definition} \em (see \cite{dnr2008})
Given two TGDs or EGDs $\alpha, \beta \in \Sigma$ we define $\alpha \prec_c \beta$ iff there exists a relational database instance $I$ and $\overline{a}, \overline{b}$ such that
(i) $I \not \models \alpha(\overline{a})$, (ii) $I \models \beta(\overline{b})$, (iii) $I \stackrel{*,\alpha, \overline{a}}{\rightarrow} J$, and (iv)~$J \not \models \beta(\overline{b})$. $\punto$
\end{definition}
 Note that in this definition we use an oblivious chase step and not a standard chase step. We give an example to illustrate this definition.

\begin{example} \em (see \cite{dnr2008} and before) Let predicate $E$ store the edge relation of a graph
and let the constraint $\alpha := E(x_1,x_2), E(x_2,x_1) \rightarrow \exists y_1, y_2 E(x_1,y_1), E(y_1,y_2), E(y_2,x_1)$ be given, stating that each node having a cycle of length $2$ also has a cycle of length $3$. A $3$-cycle can never be a $2$-cycle again, so it holds that $\alpha \not \prec_c \alpha$. $\punto$ 
\end{example}

The actual definition of c-stratification then relies, as outlined before, on the notion of weak acyclicity.

\begin{definition} \em
The c-chase graph $G_c(\Sigma)=(\Sigma,E)$ of a set of constraints $\Sigma$ contains a directed edge $(\alpha,\beta)$ between two constraints iff $\alpha \prec_c \beta$. We call $\Sigma$ c-stratified iff the constraints in every cycle of $G_c(\Sigma)$ are weakly acyclic. $\punto$
\end{definition}

\begin{figure}
\begin{center}
\begin{tabular}{c}

\begin{tikzpicture}[->,>=stealth',shorten >=1pt,auto,node distance=2cm,semithick]
\node[state] (a) [] {$\alpha_1$};
\node[state] (b) [below of=a] {$\alpha_2$};
\node[state] (c) [right of=a] {$\alpha_3$};
\node[state] (d) [below of=c] {$\alpha_4$};
\path (a) edge []          node {$$} (b)
      (a) edge [] node {$$} (c)
      (c) edge [] node {$$} (d)
      (b) edge [] node {$$} (d)
      (d) edge [] node {$$} (a);
\end{tikzpicture}	
\end{tabular}

\end{center}
\caption{Chase graph for example \ref{ex:c-strat}}
\label{fig:c-strat}
\end{figure}
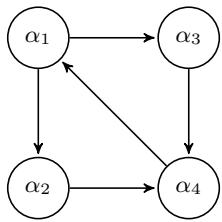

\begin{example} \em \label{ex:c-strat}
Consider the constraint set from example \ref{ex:stratfalsch} again. The problem there was that $\alpha_2$, the only TGD containing existential quantifiers, had no successor in the $\prec$-relation. However, in the c-chase graph $\alpha_2$ has a successor and indeed there is a cycle through $\alpha_2$ as witnessed by Figure \ref{fig:c-strat}. The only strongly connected component is $\Sigma$ itself, which is not weakly acyclic. So, $\Sigma$ is not c-stratified as witnessed by the non-terminating chase  sequence in example \ref{ex:stratfalsch}.
\end{example}

From the definition of $\prec_c$ it is  not immediately clear that it is decidable, however the test for membership in $\prec_c$ can be done with linear-sized databases.

\begin{proposition}
It can be decided in  $\mbox{coNP}$ whether a set of constraints is c-stratified. $\punto$
\end{proposition}

{\bf Proof Sketch.}
We start with an additional claim: let $\alpha, \beta$ be constraints. Then, the mapping $(\alpha,\beta) \mapsto \alpha \prec_c \beta ?$
can be computed by an $\mbox{NP}$-algorithm. The proof of this claim proceeds
like the proof of Theorem 3 in \cite{dnr2008}. It is enough to consider candidate
databases for $I$ of size at most $|\alpha| + |\beta|$, i.e.~unions of homomorphic
images of the premises of $\alpha$ and $\beta$ s.t.~null values occur only in
positions from $P$. Because of this claim, the c-chase graph of a
set of constraints can be computed by an $\mbox{NP}$-algorithm. To prove that $\Sigma$
is not c-stratified, guess some strongly connected component of the c-chase graph $\Sigma'$  and
verify that it is not weakly acyclic. $\qed$

And indeed c-stratification does ensure the termination of the chase independent of the data and the chase sequence as the following theorem states.

\begin{theorem} \em
Let $\Sigma$ be a fixed set of c-stratified constraints. Then, there exists a
polynomial $Q \in \mathbb{N}[X]$ such that for any database instance $I$, the length
of every chase sequence is bounded by $Q(|dom(I)|)$. $\punto$
\end{theorem}

{\bf Proof Sketch.}
Let $\Sigma$ be the set of constraints under consideration. Let $SC_1,...,SC_n$ be the strongly connected components of $G_c(\Sigma)$. We will show the following lemma.

\begin{lemma} \em
If the chase terminates data-independently for every strongly connected component of $G_c(\Sigma)$, then it terminates data-independently for $\Sigma$. $\punto$
\end{lemma}

{\bf Proof Sketch.}
Assume that we have a database instance $I_0$ such that the chase does not terminate. We will construct an infinite chase sequence that uses only constraints from some of the $C_i$. 

We have an infinite chase sequence $S = I_0 \stackrel{\alpha_1, \overline{a}_1}{\longrightarrow} I_1 \stackrel{\alpha_2, \overline{a}_2}{\longrightarrow} \ldots$.
Without loss of generality, we can assume that every constraint from $\Sigma$ fires infinitely often and that for every $j \in \mathbb{N}$ there is some $i > j$ such that $I'_{i-1} \models \alpha_{i}(\overline{a}_{i})$, where $I'_0:=I_0$, $I_{l-1} \stackrel{\alpha_l, \overline{a}_l}{\rightarrow} J_l$ for $l \neq j$ and $J_j := J_{j-1}$. Let $k \in \mathbb{N}$ arbitrary.

Consider the infinite set $W := \set{i \in \mathbb{N}_0 }{I_0 \models \alpha_i(\overline{a}_i), (\alpha_i,\overline{a}_i) \text{ appears on an arrow in }S}$. For $i \in W$ that is large enough, we find a sequence $s_i$ of constraints of length $i \geq l_i \geq 2$ such that

\begin{itemize}
	\item $s_i = \beta_{i,1} ,..., \beta_{i,l_i}$,
	\item  for every $\beta_{i,l_i}$ there is some $j$ such that $\beta_{i,l_i} = \alpha_j$,
	\item there is an instance $J_0$ such that $|J_0| \leq \sum_{j \in [l_i]} |body(\beta_{i,j})|$ and $J_0 \stackrel{\beta_{i,1}, \overline{a}_{i,1}}{\longrightarrow} J_1 \stackrel{\beta_{i,2}, \overline{a}_{i,2}}{\longrightarrow} ... \stackrel{\beta_{i,l_i}, \overline{a}_{i,l_i}}{\longrightarrow} J_{l_i+1}$, where every pair $(\beta_{i,1}, \overline{a}_{i,1})$ appears on an arrow in $S$,
	%\item $J_0 \models \beta_{i,l_i}(\overline{a}_{i,l_i})$,
	%\item there is a null value $n \in \overline{a}_{i,l_i}\backslash dom(I_0)$ in the head of $\beta_{i,l_i}(\overline{a}_{i,l_i})$,
	\item for every $j \in [l_i-1]$ it holds that $J'_{l_i-1} \models \alpha_{i,l_i}(\overline{a}_{i,l_i})$, where $J'_0:=J_0$, $J'_{l-1} \stackrel{\alpha_l, \overline{a}_l}{\rightarrow} J'_l$ for $l_i > l \neq j$ and $J'_j := J'_{j-1}$ and
	\item $l_i$ is maximal with these properties.
\end{itemize}

For every $i \in W$ large enough there is an infinite sequence of such sequences of constraints $(s_{i_j})_{j \in \mathbb{N}}$ such that $s_{i_j}$ is a subsequence of $s_{i_{j+1}}$ and $s_{i_0}=s_{i}$. Consider $C_i := \bigcup_{j \in \mathbb{N}\cup  \setone{0}} s_{i_j}$ and $C_{i,danger} := \set{\beta \in C_i}{\text{forall }k \in \mathbb{N} \text{ there is some } s_{i_j} \text{ such that } \beta \text{ appears at least }k \text{ times in }s_{i_j}}$. We find some $SC_{i'}$ ($i' \in [n]$) and $C'_i := \set{\beta_{h,l},...,\beta_{h,l_h} \in SC_{i'}}{\beta_{h,l} \in C_{i,danger}, h \in W}$
such that $C'_i \subseteq SC_{i'} \cap C_{i,danger}$. With the help of $(s_{i_j})_{j \in \mathbb{N}}$ we can show that there must be an infinite chase sequence for $C'_i$. This concludes the proof of the lemma. $\qed$\\

The proof of the theorem follows from the application of the previous lemma. The polynomial time data complexity follows from the fact that c-chase graphs bound the depth of null values (see \cite{gn2008}) data-independently. Note that by assumption, chasing with $C_i$ terminates in time $Q_i(|dom(I)|)$. $\qed$\\

\subsection{Safety}
\label{subsec:safety}

The basic idea of our new termination condition, {\it safety}, is to estimate
the set of positions where labeled nulls may be copied to and (statically)
analyze the data flow only between those positions. As a useful tool, we borrow
the notion of so-called {\it affected positions} from~\cite{cgk2008},
which is an overestimation of the positions in which a null value that
was introduced during the chase may occur.

\begin{definition} \cite{cgk2008} \em
Let $\Sigma$ be a set of TGDs. The set of affected positions $\mbox{aff}(\Sigma)$ of $\Sigma$
is defined inductively as follows. Let $\pi$ be a position in the head of an
$\alpha \in$~$\Sigma$. 
\squishlist
\item If an existentially quantified variable appears in $\pi$, then $\pi \in \mbox{aff}(\Sigma)$.
\item If the same universally quantified variable $X$ appears both in position $\pi$
and only in affected positions in the body of $\alpha$, then
$\pi \in \mbox{aff}(\Sigma)$. $\punto$
\squishend
\end{definition}

Although we borrow this definition from~\cite{cgk2008}, our focus is different.
We use affected positions to extend known classes of constraints for which
the chase terminates, whereas~\cite{cgk2008}~investigates query answering in
cases the chase may not terminate. Our work neither subsumes~\cite{cgk2008}
nor the other way around.

\begin{figure}[t]
\begin{center}
	\begin{tabular}[t]{cc|cc} 
\includegraphics[width=3cm]{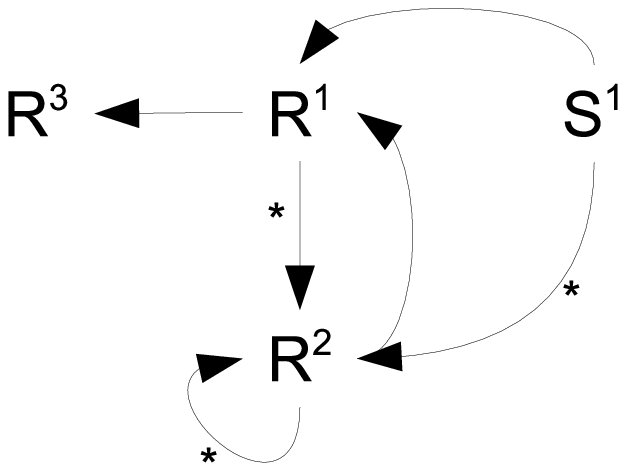}
		& \hspace{0.2cm} & \hspace{0.2cm} &
	\raisebox{1cm}{\includegraphics[width=0.5cm]{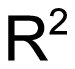}}\\
	\end{tabular}
\end{center}
%\vspace{-0.3cm}
\caption{Left: Dependency graph. Right: Corresponding propagation graph (it has no edges).}
%\vspace{-0.4cm}
\label{tab:notweak--safe}
\end{figure}

We motivate the safety termination condition using the single constraint
$\beta := R(x_1,x_2,x_3), S(x_2) \rightarrow \exists y R(x_2,y,x_1)$.
The dependency graph of constraint set $\{ \beta \} $ is shown
in Figure~\ref{tab:notweak--safe} (left). As can be seen,
there is a cycle going through a special edge, so the set 
is not weakly acyclic. We next study the affected positions in $\beta$:

\begin{example} \label{ex:safe} \em
Let us consider the constraint set $\Sigma := \{ \beta \}$.
Clearly, position $R^2$ is affected because it contains an existentially
quantified variable. $S^1$ is not affected because $S$ is not modified
when chasing with the single constraint $\beta$. Finally, we observe that
also $R^1$ is not affected because $x_2$ occurs not only in $R^2$
but also in $S^1$, which is not an affected position.
We conclude that position~$R^2$ is the
only affected position in constraint set $\Sigma$.$\punto$
\end{example}

We now argue that for constraint~$\beta$ a cascading of
fresh labeled nulls cannot occur, i.e.~no fresh labeled null can repeatedly
create new labeled nulls in position $R^2$ while copying itself to position~$R^1$.
The reason is that $\beta$ cannot be violated with a fresh labeled
null in $R^2$, i.e.~if $R(a_1,a_2,a_3)$ and  $S(a_2)$ hold,
but $\exists y R(a_2,y,a_1)$ does not, then $a_2$ is never a newly created
labeled null. This is due to the fact that $a_2$ also occurs in~$S^1$,
but $S^1$ is not an affected position. Hence, the chase sequence
always terminates.  We will later see that this is not a mere coincidence:
the constraint is safe.

Like in the case of weak acyclicity, we define the safety condition with
the help of the absence of  cycles containing special edges in some graph,
called propagation graph.

\begin{definition} \em
Given a set of TGDs $\Sigma$, the propagation graph $\mbox{prop}(\Sigma):=(\mbox{aff}(\Sigma),E)$ is the directed graph defined as follows. There are two kinds of edges in $E$. Add them as follows: for every TGD 
$\forall \overline{x}(\phi(\overline{x}) \rightarrow \exists \overline{y} \psi(\overline{x},\overline{y})) \in \Sigma$
 and for every $x$ in $\overline{x}$ that occurs in $\psi$ and every occurrence of $x$ in $\phi$ in position $\pi_1$
\squishlist
	\item if $x$ occurs only in affected positions in $\phi$ then, for every occurrence of $x$ in $\psi$ in position $\pi_2$, add an edge $\pi_1 \rightarrow \pi_2$ (if it does not already exist).
	\item if $x$ occurs only in affected positions in $\phi$ then, for every existentially quantified variable $y$ and for every occurrence of $y$ in a position $\pi_2$, add a {\it special} edge $\pi_1 \stackrel{*}{\rightarrow} \pi_2$ (if it does not already exist).$\punto$
\squishend
\label{def:prograph}
\end{definition}

As an improvement over weak acyclicity, in the propagation graph 
we do not supervise the whole data flow but only the flow of labeled nulls
that might be introduced at runtime. Consequently, the graph contains edges only for 
null values that stem exclusively from affected positions.
We now can easily define the safety condition on top of the propagation graph.

\begin{definition} \em
A set $\Sigma$ of constraints is called \textit{safe} iff $\mbox{prop}(\Sigma)$
has no cycles going through a special edge. $\punto$
\label{def:safety}
\end{definition}

\begin{example} \em \label{ex:safe2}
Consider the constraint $\beta$ from Example~\ref{ex:safe}.
Its dependency graph is depicted in Figure \ref{tab:notweak--safe} on the
left side and its propagation graph on the right side. The latter contains
only the affected position $R^2$ (and no edges). From
Definitions~\ref{def:wa} and~\ref{def:safety} it follows that 
$\beta$ is safe, but not weakly acyclic. $\punto$
\end{example}

The intuition of safety is that we forbid an unrestricted cascading
of null values, i.e.~with the help of the propagation graph we
impose a partial order on the affected positions such that any
newly introduced null value can only be created in a position that
has a higher rank in that partial order in comparison to null values
that may occur in the body of a TGD. To state this more precisely,
assume that a TGD of the form
$\forall \overline{x}(\phi(\overline{x}) \rightarrow \exists \overline{y} \psi(\overline{x},\overline{y}))$
is violated. Then,
$I \models \phi(\overline{a})$ and $I \not \models \exists \overline{y} \psi(\overline{a},\overline{y}))$
must hold. The safety condition ensures that any position in the body
that contains a newly created labeled null from $\overline{a}$ 
and occurs in the head of the TGD has a strictly lower rank in
our partial order than any position in which some element from $\overline{y}$
occurs. The main difference compared to weak acyclicity is that, in safety,
we look in a refined way (cf.~affected positions) on positions where labeled
nulls can be propagated to.

It is easy to see that, if a constraint set $\Sigma$ is safe, then
every subset of $\Sigma$ is safe, too. Furthermore, we note that,
given a set of constraints, it can be decided in polynomial time if
it is safe or not. In the following theorem we relate safety to the
previous termination conditions weak acyclicity and stratification.
In particular, the theorem clarifies the observation from
Example~\ref{ex:safe2}, where we could observe that the propagation
graph is a subgraph of the dependency graph. This is not a mere coincidence:

\begin{theorem} \label{rel-safety} \em
Let $\Sigma$ be a set of constraints.
\squishlist
\item The graph $\mbox{prop}(\Sigma)$ is a subgraph of $\mbox{dep}(\Sigma)$.
\item If $\Sigma$ is weakly acyclic, then it is also safe. 
\item There is some $\Sigma$ that is safe, but not c-stratified and vice versa. $\punto$
\squishend
\end{theorem}

{\bf Proof Sketch.}
(a) The set of vertices from $\mbox{prop}(\Sigma)$ is contained in the
set of vertices of $\mbox{dep}(\Sigma)$. In order to add an edge to
$\mbox{prop}(\Sigma)$ stronger prerequisites must be fulfilled than
in the construction of $\mbox{dep}(\Sigma)$. Therefore $\mbox{prop}(\Sigma)$
is a subgraph of $\mbox{dep}(\Sigma)$. (b)~If $\mbox{dep}(\Sigma)$ does
not have a cycle through a special edge, then $\mbox{prop}(\Sigma)$ cannot have.
(c) Let $\alpha := S(x_2,x_3), R(x_1,x_2,x_3) \rightarrow \exists y R(x_2,y,x_1)$ and $\beta := R(x_1,x_2,x_3) \rightarrow S(x_1,x_3)$. It can be seen that $\alpha \prec \beta$ and $\beta \prec \alpha$. Together with the fact that $\setone{\alpha, \beta}$ is not weakly acyclic it follows that $\setone{\alpha, \beta}$ is not stratified. However, $\setone{\alpha, \beta}$ is safe. 
Let $\gamma$ $:=$ $T(x_1,x_2),$ $T(x_2,x_1)$ $\rightarrow$ $\exists$ $y_1,y_2$ $T(x_1,y_1),$ $T(y_1,y_2),$ $T(y_2,x_1)$ 	(see \cite{dnr2008}). It was argued in \cite{dnr2008} that $\setone{\gamma}$ is stratified. However, it is not safe because both $T^1$ and $T^2$ are affected. Therefore we have that $\mbox{dep}(\setone{\gamma})=\mbox{prop}(\setone{\gamma})$ and it was argued in \cite{dnr2008} that it is not weakly acyclic. $\qed$

Like stratification and weak acyclicity, safety guarantees the termination
of the chase in polynomial time data complexity, i.e. the set of constraints
is fixed and the number of chase steps is polynomial in the number of
distinct values in the input database instance:

\begin{theorem} \label{SAFEMAIN} \em
Let $\Sigma$ be a fixed set of safe constraints. Then, there exists a
polynomial $Q \in \mathbb{N}[X]$ such that for any database instance $I$, the length
of every chase sequence is bounded by $Q(|dom(I)|)$. $\punto$
\end{theorem}

{\bf Proof Sketch.}
First we introduce some additional notation. We denote constraints in the form 
$\phi(\overline{x_1},\overline{x_2},\overline{u}) \rightarrow \exists \overline{y} \psi(\overline{x_1},\overline{x_2},\overline{y})$, where $\overline{x_1},\overline{x_2},\overline{u}$ are all the universally quantified variables and 
\squishlist
	\item $\overline{u}$ are those variables that do not occur in the head,
	\item every element in $\overline{x_1}$ occurs in a non-affected position in the body, and
	\item every element in $\overline{x_2}$ occurs only in affected positions in the body.
\squishend

The proof is inspired by the proof of Theorem 3.8 in \cite{fkmp2005}, especially the notation and some introductory definitions are taken from there. In a first step we will give the proof for TGDs only, i.e.~we do not consider EGDs. Later, we will see what changes when we add EGDs again.

Note that $\Sigma$ is fixed. Let $(V,E)$ be the propagation graph $\mbox{prop}(\Sigma)$. For every position $\pi \in V$ an incoming path is a, possibly infinite, path ending in $\pi$. We denote by $rank(\pi)$ the maximum number of special edges over all incoming paths. It holds that $rank(\pi) < \infty$
 because $\mbox{prop}(\Sigma)$ contains no cycles through a special edge. Define $r := max\set{rank(\pi)}{\pi \in V}$ and $p := |V|$. It is easily verified that $r \leq p$, thus $r$ is bounded by a constant. This allows us to partition the positions into sets $N_0,...,N_p$ such that $N_i$ contains exactly those positions $\pi$ with $rank(\pi) = i$. Let $n$ be the number of values in $I$. We define $dom(\Sigma)$ as the set of constants in $\Sigma$.

Choose some $\alpha := \phi(\overline{x_1},\overline{x_2},\overline{u}) \rightarrow \exists \overline{y} \psi(\overline{x_1},\overline{x_2},\overline{y}) \in \Sigma$. Let $I \rightarrow \ldots \rightarrow \overline{G} \stackrel{\alpha,\overline{a}_1 \text{ }\overline{a}_2 \text{ } \overline{b}}{\longrightarrow} G'$ and let $\overline{c}$ be the newly created null values in the step from $\overline{G}$ to $G'$. Then

\begin{enumerate}
  \item newly introduced labeled nulls occur only in affected positions,
	\item $\overline{a_1} \subseteq dom(I) \cup dom(\Sigma)$ and
	\item for every labeled null $Y \in \overline{a_2}$ that occurs in $\pi$ in $\phi$ and every $c \in \overline{c}$ that occurs in $\rho$ in $\psi$ it holds that $rank(\pi) < rank(\rho)$.
\end{enumerate}

This intermediate claim is easily proved by induction on the length of the chase sequence. Now we show by induction on $i$ that the number of values that can occur in any position in $N_i$ in $G'$ is bounded by some polynomial $Q_i$ in $n$ that depends only on $i$ (and, of course, $\Sigma$). As $i \leq r \leq p$, this implies the theorem's statement because the maximal arity $ar(\mathcal{R})$ of a relation is fixed. We denote by ${\it body}(\Sigma)$ the number of characters of
the largest body of all constraints in $\Sigma$.\\

\underline{Case 1: $i=0$.}  We claim that $Q_0(n)$:=$n + |\Sigma| \cdot n^{ar(\mathcal{R}) \cdot body(\Sigma)}$
 is sufficient for our needs. We consider a position $\pi \in N_0$ and an arbitrary TGD from $\Sigma$ such that $\pi$ occurs in the head of $\alpha$. For simplicity we assume that it has the syntactic form of $\alpha$. In case that there is a universally quantified variable in $\pi$, there can occur at most $n$ distinct elements in $\pi$. Therefore, we assume that some existentially quantified variable occurs in $\pi$ in $\psi$. Note that as $i=0$ it must hold that $|x_2|=0$. Every value in $I$ can occur in $\pi$. But how many labeled nulls can be newly created in $\pi$? For every choice of $\overline{a_1} \subseteq dom(\overline{G})$ such that 
$\overline{G} \models \phi(\overline{a_1},\lambda,\overline{b})$ and $\overline{G} \nvDash \exists \overline{y} \psi(\overline{a_1},\lambda,\overline{y})$ at most one labeled null can be added to $\pi$ by $\alpha$. Note that in this case it holds that $\overline{a_1} \subseteq dom(I)$ due to (1). So, there are at most 
$n^{ar(\mathcal{R}) \cdot body(\Sigma)}$ such choices. Over all TGDs at most $|\Sigma| \cdot n^{ar(\mathcal{R}) \cdot body(\Sigma)}$ labeled nulls are created in $\pi$.\\

\underline{Case 2: $i \rightarrow i+1$.}  We claim that $Q_{i+1}(n) := \sum_{j=0}^{i} Q_i(n) + |\Sigma| \cdot (\sum_{j=0}^{i} Q_i(n))^{ar(\mathcal{R}) \cdot body(\Sigma)}$ is such a polynomial. Consider the fixed TGD $\alpha$. Let $\pi \in N_{i+1}$. Values in $\pi$ may be either copied from a position in $N_0 \cup ... \cup N_i$ or may be a new labeled null. Therefore w.l.o.g.~we assume that some existentially quantified variable occurs in $\pi$ in $\psi$. In case a TGD, say $\alpha$, is violated in $G'$ there must exist $\overline{a_1}, \overline{a_2} \subseteq dom_{G'}(N_0,...,N_i)$ and $\overline{b} \subseteq dom(G')$ such that $G' \models \phi(\overline{a_1},\overline{a_2},\overline{b})$, but $G' \nvDash \exists \overline{y} \psi(\overline{a_1},\overline{a_2},\overline{y})$.  If newly introduced labeled null occurs in $\overline{a_2}$, say in some position $\rho$, then $\rho \in \bigcup_{j=0}^{i} N_j$. As there are at most 
$(\sum_{j=0}^{i} Q_i(n))^{ar(\mathcal{R}) \cdot body(\Sigma)}$ many such choices for $\overline{a_1}, \overline{a_2}$, at most $(\sum_{j=0}^{i} Q_i(n))^{ar(\mathcal{R}) \cdot body(\Sigma)}$ many labeled nulls can be newly created in $\pi$.\\

When we allow EGDs among our constraints, we have that the number of values that can occur in any position in $N_i$ in $G'$ can be bounded by the same polynomial $Q_i$ because equating labeled nulls does not increase the number of labeled nulls and the fact that EGDs preserve valid existential conclusions of TGDs.  $\qed$

\subsection{Inductive Restriction}

In this section we generalize the method that lifts weak acyclicity to stratification
from~\cite{dnr2008} with the help of so-called {\it restriction systems}. The chase
graph from~\cite{dnr2008} will be a special case of such a restriction system. With
the help of restriction systems we then define a new sufficient termination condition
called {\it inductive restriction}, whose main idea is to decompose a given constraint
set into smaller subsets (in a more refined way than stratification). We then use the
safety condition from before to check the termination of every subset and,
whenever all subsets are safe, the termination for the full constraint set can be
guaranteed. Ultimately, we show that inductive restriction
(like all the classes discussed before) guarantees chase termination in
polynomial-time data complexity. This section also lays the
foundations for the $T$-hierarchy (cf.~Figure~\ref{fig:survey}), which will be
defined subsequently in Section~\ref{sec:t}. 
We motivate our study with a constraint set that is neither safe nor stratified.

%Let $\alpha := S(x_2,x_3), R(x_1,x_2,x_3) \rightarrow \exists y R(x_2,y,x_1)$
%and $\beta := R(x_1,x_2,x_3) \rightarrow S(x_1,x_3)$. It can be seen that $\alpha \prec \beta$
%and $\beta \prec \alpha$. Further, $\setone{\alpha, \beta}$ is not weakly
%acyclic, so it follows that $\setone{\alpha, \beta}$ is not stratified. Still, the chase will
%always terminate: A firing of $\alpha$ may cause a null value to appear in position $R^2$,
%but a firing of $\beta$ will never introduce null values in the head of $\beta$ although
%$\beta \prec \alpha$ holds. 

\begin{example} \em 
Let predicate E($x$,$y$) store graph edges and predicate S($x$) store some nodes.
The constraints set $\Sigma = \{ \alpha_1, \alpha_2 \}$ with
$\alpha_1 := S(x), E(x,y) \rightarrow E(y,x)$ and
$\alpha_2 := S(x), E(x,y) \rightarrow \exists z E(y,z), E(z,x)$
assert that all nodes in S have cycles of length $2$ and $3$, respectively.
It holds that $\mbox{aff}(\Sigma)$ = $\{E^1,E^2\}$ and it is easy to verify that~$\Sigma$
is neither safe nor stratified. In particular, it we observe that 
$\alpha_1 \prec \alpha_2$ and $\alpha_2 \prec \alpha_1$. $\punto$
\label{ex:notsafestratified1}
\end{example}

The first task in our formalization is a refinement of relation
$\prec$ from~\cite{dnr2008}. This refinement will helps us to detect
if during the chase null values might be copied to the head of
some constraint. To simplify the definition, we introduce the
notion of $\mbox{null-pos}$: 

\begin{definition}  \em
Let $\Sigma$ be a set of constraints, $I$ be a fixed database instance
and $A \subseteq \Delta_{null}$. Then, we define $\mbox{null-pos}(A,I)$ as
$\{ \pi \in pos(\Sigma) \mid a \in A, a \text{ occurs in position } \pi \text{ in }I \}$. $\punto$
\end{definition}

Informally spoken, $\mbox{null-pos}(A,I)$ is the set of positions in $I$
in which the elements (i.e., labeled nulls) from $A$ occur. We are now
ready to define the refinement of relation $\prec$:

\begin{definition} \label{def-verf} \em
Let $\Sigma$ be a set of constraints and $P \subseteq pos(\Sigma)$. For all $\alpha, \beta \in \Sigma$, we define $\alpha \prec_{P} \beta$ iff there are tuples $\overline{a}, \overline{b}$ and a database instance $I_0$ such that 
\squishlist 
\item $I_0 \stackrel{*,\alpha, \overline{a}}{\rightarrow} I_1$, 
\item $I_1 \not \models \beta(\overline{b})$,
\item there is $n \in \overline{b} \cap \Delta_{null}$ in the head of $\beta(\overline{b})$ such that $\mbox{null-pos}(\setone{n},I_0) \subseteq P$, and
\item $I_0 \models \beta(\overline{b})$. $\punto$
\squishend
\end{definition}

The refinement of $\prec$ forms the basis for the notion of a so-called
restriction system, which is a strict generalization of the chase graph
introduced in~\cite{dnr2008} and will serve as a central tool in our work.
The two definitions below formalize restriction systems.

\begin{definition} \em
For any set of positions $P$ and a TGD $\alpha$ let $\mbox{aff-cl}(\alpha,P)$ be the set of positions $\pi$ from the head of $\alpha$ such that 

\squishlist
	\item for every universally quantified variable $x$ in $\pi$: $x$ occurs in the body of $\alpha$ only in positions from $P$ or
	\item $\pi$ contains an existentially quantified variable. $\punto$
\squishend
\end{definition}

\begin{definition} \label{rest} \em
A {\it $2$-restriction system\footnote{In \cite{msl2009,msl2_2009,sml2008} the notion of a $2$-restriction system was simply called restriction system and was defined slightly different there.}} is a pair $(G'(\Sigma),f)$, where
$G'(\Sigma) := (\Sigma,E)$ is a directed graph and
$f \subseteq pos(\Sigma)$ such that
\squishlist
	\item forall TGDs $\alpha$ and forall $(\alpha,\beta) \in E$: $\mbox{aff-cl}(\alpha,f) \cap pos(\Sigma) \subseteq f$ and $\mbox{aff-cl}(\beta,f) \cap pos(\Sigma) \subseteq f$,
	%	\item forall EGDs $\alpha$ and forall $(\alpha,\beta) \in E$: $f(\alpha) \subseteq f(\beta)$, and
	\item forall $\alpha, \beta \in \Sigma$: $\alpha \prec_{f} \beta \implies (\alpha,\beta) \in E$. 
\squishend

A $2$-restriction system is {\it minimal} if it is obtained from
($(\Sigma,\emptyset)$,$\emptyset$) by
a repeated application of the constraints from bullets one to three (until all
constraints hold) s.t., in case of the first , $f$ is extended only by those positions that are required to satisfy
the condition. $\punto$
\end{definition}

We illustrate this definition by two examples. The first one also shows that
restriction systems always exist.

\begin{example} \label{alle} \em
Let $\Sigma$ a set of constraints. Then, $(G(\Sigma),pos(\Sigma))$, is a $2$-restriction system for constraint set $\Sigma$. $\punto$
\end{example}

\begin{example} \em
Consider $\Sigma$ from Example~\ref{ex:notsafestratified1}. The minimal
$2$-restriction system for $\Sigma$ is
G'($\Sigma$):=($\Sigma$,\{($\alpha_2$,$\alpha_1$)\}) with
$f := \{E^1,E^2\}$; in particular, 
$\alpha_1 \not \prec_{f)} \alpha_1$,
$\alpha_1 \not \prec_{f} \alpha_2$,
$\alpha_2 \prec_{f} \alpha_1$, and
$\alpha_2 \not \prec_{f} \alpha_2$ hold.
$\punto$
\label{ex:sr1}
\end{example}

Restriction systems are useful tools to define new classes of constraints
that guarantee chase termination. To give an example, one can show that the
chase with a constraint set $\Sigma$ terminates for every database instance
if every strongly connected component of its minimal $2$-restriction system
is safe. We refer the interested reader to~\cite{msl2009} for details,
where this class was formally introduced under the name {\it safe restriction}.
Note that the constraint set $\Sigma$ from Example~\ref{ex:notsafestratified1}
falls into the class of safely restricted constraints, because
its minimal $2$-restriction system (given in Example~\ref{ex:sr1})
contains no strongly connected component. In this work, we skip the
formal definition of safe restriction, but instead go one step further
and define a termination condition called {\it inductive restriction}, which
further generalizes safe restriction. The following example
provides a constraint set that is not safely restricted but, as we
shall see later, falls into the class of inductively restricted
constraints.

%\begin{definition} \label{def-rest} \em
%$\Sigma$ is called safely restricted if and only if there is a restriction system $(G'(\Sigma),f)$ for $\Sigma$ such that every strongly connected component in $G'(\Sigma)$ is safe. $\punto$
%\end{definition}
%
%The next theorem shows that safe restriction strictly extends the notion of stratification and safety.
%
%\begin{theorem} \label{rest-safe-weak} \em If $\Sigma$ is stratified or safe, then it is also safely restricted. There is some $\Sigma$ that is safely restricted but neither safe nor stratified. $\punto$
%\end{theorem}
%
%Definition \ref{def-rest} implies that safely restricted constraints can be recognized by a $\Sigma_2^P$-algorithm. However, with the help of a canonical restriction system, we can show that safe restriction can be decided in $\textsc{coNP}$ (like stratification).
%
%\begin{theorem} \label{rest-comp} \em
%Given constraint set $\Sigma$ it can be checked by a $\textsc{coNP}$-algorithm whether $\Sigma$ is safely restricted. $\punto$
%\end{theorem}
%
%The next theorem is the main contribution of this section. It states that the chase will always terminate in polynomial time data complexity for safely restricted constraints. 
%
%\begin{theorem} \label{chase-main} \em
%Let $\Sigma$ be a fixed set of safely restricted constraints. Then, there exists a polynomial $Q \in \mathbb{N}[X]$ such that for any database instance $I$, the length of every chase sequence is bounded by $Q(||I||)$, where $||I||$ is the number of distinct values in $I$. $\punto$
%\end{theorem}

\begin{example} \em
We extend the constraint set from Example~\ref{ex:sr1} to
$\Sigma' := \Sigma \cup \{ \alpha_3 \}$, where $\alpha_3 := \exists x, y S(x), E(x,y)$.
Then G'($\Sigma'$):=($\Sigma'$,\{($\alpha_1,\alpha_2$),($\alpha_2$,$\alpha_1$),($\alpha_3$,$\alpha_1$),($\alpha_3$,$\alpha_2$)\})
with $f\{E^1,E^2,S^1\}$ is the minimal $2$-restriction system. It contains the strongly
connected component \{$\alpha_1$,$\alpha_2$\}. 
Note that $\Sigma'$ is neither safe, nor stratified, nor safely restricted.
Hence, using the sufficient termination conditions discussed so far no chase 
termination guarantees can be made for $\Sigma'$.$\punto$
\label{ex:ir1}
\end{example}

\begin{figure}[t]
\centering
\begin{boxedminipage}{6.5cm}
\begin{tabbing}
l \= lllll \= lllllll \= lllllllll \= \kill
{\bf part}($\Sigma$: Set of TDGs and EGDs, $k$: not equal to $1$) \{\\
\>1:\>compute the strongly connected components (as\\
\>\> sets of constraints) $C_1$, $\dots$, $C_n$ of the minimal\\
\>\> $k$-restriction system of $\Sigma$;\\
\>2:\>$D \leftarrow \emptyset$\\
\>3:\>{\bf if}\ \  (n == 1)\ \  {\bf then}\ \ \\
\>4:\>\> {\bf if}\ \ ($C_1 \not = \Sigma$)\ \ {\bf then}\ \ \\
\>5:\>\>\> return {\bf part}($C_1$,$k$);\\
\>6:\>\> {\bf endif}\\
\>7:\>\> return $\setone{\Sigma}$;\\
\>8:\> {\bf endif}\\\>6:\>{\bf for}\ \  i=1 to n\ \  {\bf do}\ \ \\
\>9:\>\> $D \leftarrow D\ \cup$ {\bf part}($C_i$,$k$);\ \ \\
\>10:\>{\bf endfor}\\
\>11:\>return $D$; \}
\end{tabbing}
\end{boxedminipage}
%\vspace{-0.3cm}
\caption{Algorithm to compute subsets of $\Sigma$.}
%\vspace{-0.4cm}
\label{fig:algopart}
\end{figure}

Intuitively, in the example above the constraint
$\alpha_3$ ``infects'' position $S^1$ in the $2$-restriction system. Still,
null values cannot be repeatedly created in $S^1$: $\alpha_3$ fires at most once,
so it does not affect chase termination. Our novel termination
condition resolves such situations by recursively computing the minimal $2$-restriction
systems of the strongly connected components. We formalize this computation
in Algorithm~1, called ${\it part}(\Sigma,2)$ and define the class of inductively
restricted constraint sets by help of this algorithm.

\begin{definition} \em
Let $\Sigma$ be a set of constraints. We call $\Sigma$ {\it inductively
restricted} iff every $\Sigma' \in {\it part}(\Sigma,2)$ is safe.$\punto$
\end{definition}

Compared to stratification, inductive restriction does
not increase the complexity of the recognition problem:

\begin{lemma} \label{relindsafe} \em Let $\Sigma$ be a set of constraints.
The recognition problem for inductive restriction is in $\mbox{coNP}$. $\punto$
\label{lemma:ir}
\end{lemma}

{\bf Proof Sketch.}
We start with an additional claim: let $P$ be a set of positions and $\alpha, \beta$
constraints. Then, the mapping $(P,\alpha,\beta) \mapsto \alpha \prec_P \beta ?$
can be computed by an $\mbox{NP}$-algorithm. The proof of this claim proceeds
like the proof of Theorem 3 in \cite{dnr2008}. It is enough to consider candidate
databases for $I_0$ of size at most $|\alpha| + |\beta|$, i.e.~unions of homomorphic
images of the premises of $\alpha$ and $\beta$ s.t.~null values occur only in
positions from $P$. Because of this claim, the minimal $2$-restriction system of a
set of constraints can be computed by an $\mbox{NP}$-algorithm (only polynomially
many steps must be performed to reach the fixedpoint). Computing $part(\Sigma,2)$
can also be done in non-deterministic polynomial time. To prove that $\Sigma$
is not inductively restricted, guess some $\Sigma' \in part(\Sigma,2)$ and
verify that it is not safe. $\qed$
\medskip

We give an example for an inductively restricted constraint set, which
-- as argued in Example~\ref{ex:ir1} -- is neither safe nor
stratified.

\begin{example} \label{bulletzwo} \em
Referring back to Example~\ref{ex:ir1}, we have seen that the minimal $2$-restriction system of $\Sigma'$ contains the only strongly connected component \{$\alpha_1$,$\alpha_2$\}, which by Example~\ref{ex:notsafestratified1} is not safe. Therefore we compute the minimal $2$-restriction system of \{$\alpha_1$,$\alpha_2$\} and see that it does not contain a cycle. This argumentation proves that
${\it part}(\Sigma',2) = \emptyset$, so we conclude that constraint set
$\Sigma'$ is inductively restricted. $\punto$
\label{ex:ir2}
\end{example}

As depicted in Figure~\ref{fig:survey}, the inductive restriction condition
generalizes both safety and weak acyclicity. The following proposition formally states these
results and shows that the respective inclusion relationships are proper. Please note that, as we will show later inductive restriction ensures chase termination independent of the database and the chase sequence, therefore it cannot extend stratification, see example \ref{ex:stratfalsch}.

\begin{proposition} \em The following claims hold. 
\squishlist
\item If $\Sigma$ is safe, then it is inductively restricted. 
\item There is some $\Sigma$ that is stratified, but not inductively restricted.
\item There is some $\Sigma$ that is inductively restricted, but neither safe nor c-stratified. $\punto$
\squishend
\end{proposition}

{\bf Proof Sketch.}
We start with bullet one.
It follows from the fact that every subset of a safe constraint set is safe.
Bullet two follows from example \ref{ex:stratfalsch}.
Finally, bullet three is proven by the constraint set from
Examples~\ref{ex:ir1} and \ref{ex:ir2}.$\qed$

\medskip

The next theorem gives the main result concerning inductive restriction, showing that
it guarantees chase termination in polynomial time data complexity. We refer the interested
reader to theorem \ref{main-tk} for a formal proof of this theorem.

\begin{theorem} \label{IND-MAIN} \em
Let $\Sigma$ be a fixed set of inductively restricted constraints. Then, there exists a
polynomial $Q \in \mathbb{N}[X]$ such that for any database instance $I$, the length
of every chase sequence is bounded by $Q(|dom(I)|)$. $\punto$
\end{theorem}

We conclude with the remark that our motivating constraint set from
Figure~\ref{fig:schemaandconstraints1} is not inductively restricted:
the constraint $\alpha$ can cause itself to fire, so its minimal $2$-restriction
system contains an edge from $\alpha$ to $\alpha$, which forms a strongly connected
component; further, $\alpha$ is not safe. To show that the chase with $\alpha$
terminates, we need weaker termination conditions than inductive restriction.

\subsection{The T-Hierarchy}
\label{sec:t}

This section introduces the $T$-hierarchy, which is our main result regarding data-independent
chase termination. Its lowest level, $T[2]$, corresponds to inductive restriction. Every
level in the hierarchy is decidable and contains all lower levels. As we shall see,
also the constraint from Figure~\ref{fig:schemaandconstraints1} is a member of some
level in this hierarchy.
In the course of this section we leave out some proofs for space limitations,
referring the interested reader to the technical report~\cite{msl3_2009}.
We start by defining the $k$-ary relation $\prec_{k,P}$ which is a generalization of $\prec_P$.
The definition naturally extends the $\prec_P$ relation to a fixed number
$k$ of constraints.

\begin{definition}  \em
Let $k \geq 2$, $\Sigma$ a set of constraints and $P \subseteq pos(\Sigma)$. For all $\alpha_1,..., \alpha_k \in \Sigma$, we define $\prec_{k,P}(\alpha_1,...,\alpha_k)$ iff there are tuples $\overline{a}_1,...,\overline{a}_k$ and a database instance $I_0$ such that 

\squishlist
\item for all $i \in [k-1]$ it holds that $I_{i-1} \stackrel{*,\alpha_i, \overline{a}_i}{\rightarrow} I_i$,
\item $I_{k-1} \not \models \alpha_k(\overline{a}_k)$,
%\item for all $i \in [k-1]$ there is $j \in [k] \backslash [i]$: $I'_{j-1} \models \alpha_j(\overline{a}_j)$, where $I'_0 := I_0$, $I_{l-1} \stackrel{\alpha_l, \overline{a}_l}{\rightarrow} I_l$ for $l \neq i$ and $I'_i := I'_{i-1}$,
\item there is $n \in \overline{a}_k \cap \Delta_{null}$ in the head of $\alpha_k(\overline{a}_k)$ such that $\mbox{null-pos}(\setone{n},I_0) \subseteq P$,
\item $I_0 \models \alpha_k(\overline{a}_k)$, and 
\item for every $i \in [k-1]$ it holds that $J_{k-1}$ is defined and $J_{k-1} \models \alpha_k(\overline{a}_k)$, where $J_0:=I_0$, $J_{l-1} \stackrel{*,\alpha_l, \overline{a}_l}{\rightarrow} J_l$ for $k > l \neq i$ and $J_i := J_{i-1}$. $\punto$
\squishend
\end{definition}

Note that $\prec_{2,P}$ corresponds exactly to $\prec_P$ introduced in
Definition~\ref{def-verf}. It can be shown that, for a fixed value of $k$,
membership in this relation is decidable in $\mbox{NP}$: 

\begin{proposition} \em \label{lemma2} \label{LEMMA2}
Let $k \geq 2$ be fixed. Then, there exists a $\mbox{NP}$-algorithm that decides for every set of positions $P$ and every $\alpha_1,...,\alpha_k \in \Sigma$ whether $\prec_{k,P}(\alpha_1,...,\alpha_k)$ holds. $\punto$
\end{proposition}

{\bf Proof:} Let $k \geq 2$ be fixed and $\alpha_1,...,\alpha_k$ be TGDs and EGDs. Assume $\prec_{k,P}(\alpha_1,...,\alpha_k)$. Choose a database instance $I_0$ and sequences $\overline{a}_1,...,\overline{a}_k$ such that the definition of $\prec_{k,P}(\alpha_1,...,\alpha_k)$ holds.  For all $i \in [k-1]$ set $I_{i-1} \stackrel{*,\alpha_i, \overline{a}_i}{\rightarrow} I_i$ and $I_{k-1} \stackrel{*,\alpha_k, \overline{a}_k}{\rightarrow} I_k$. There is a sequence of homomorphisms $h_1,...,h_k$ such that $h_i: body(\alpha_i) \rightarrow I_{i-1}$ for all $i \in [k]$. Let $J_{0} \subseteq I_{0}$ be the minimal subinstance (with respect to set cardinality) such that for all $i \in [k]$ $J_{i-1} \stackrel{\alpha_i, \overline{a}_i}{\rightarrow} J_i$ and $J_i \subseteq I_i$. Then $J_0$ and $\overline{a}_1,...,\overline{a}_k$ also satisfy the conditions from the definition of $\prec_{k,P}(\alpha_1,...,\alpha_k)$. Furthermore. it must hold that $|J_0| \leq \sum_{i \in [k]} (|body(\alpha_i)| + |head(\alpha_i)|) \leq \sum_{i \in [k]} |\alpha_i|$, where $|\alpha_i|$ denotes the length of the sequence of symbols of the formula $\alpha_i$. So only finitely many candidate databases have to be examined, which completes the proof. $\qed$

We next use the relation $\prec_{k,P}$ to define $k$-restriction systems, which naturally
generalize the $2$-restriction systems defined over relation
$\prec_P$ (cf.~Definition~\ref{rest}).

\begin{definition} \em
Let $k \in \mathbb{N}_{>1}$. A $k$-restriction system $G'_{k}(\Sigma)$ is a pair $(G',f)$, where $G'=(\Sigma,E)$ is a graph and $f \subseteq pos(\Sigma)$ such that

\squishlist
	\item forall TGDs $\alpha$ and forall $(\alpha,\beta) \in E$: $\mbox{aff-cl}(\alpha,f) \subseteq f$ and
		%\item forall EGDs $\alpha$ and forall $(\alpha,\beta) \in E$: $f(\alpha) \subseteq f(\beta)$, and	
	\item forall $\alpha_1,..., \alpha_{k} \in \Sigma$: $\prec_{k,f}(\alpha_1,...,\alpha_{k})$ then $(\alpha_1,\alpha_2),...,(\alpha_{k-1},\alpha_{k}) \in E$. 
\squishend

A $k$-restriction system is {\it minimal} if it is obtained from
($(\Sigma,\emptyset),\emptyset$) by
a repeated application of the constraints from bullets one and (until all
constraints hold) such that, in case of the first bullet, $f$  is extended only by those positions that are required to satisfy
the condition. In case the second bullet is applied, $E$ is extended. $\punto$
\end{definition}

Note that for $k=2$ this definition corresponds exactly to the
definition of $2$-restriction systems used to define inductive
restriction. Like $2$-restriction systems, minimal
$k$-restriction systems are unique and can be computed by a
$\mbox{coNP}$-algorithm:

\begin{proposition} \label{krest}  \label{KREST} \em
Let $k \geq 2$ be fixed and $\Sigma$ a set of constraints. The minimal
$k$-restriction system for $\Sigma$ is unique and can be computed
by a $\mbox{NP}$-algorithm. $\punto$
\end{proposition}

{\bf Proof:} Uniqueness follows directly from the definition: the computation is monotone and bounded. The computation takes polynomially many steps and each step requires at most one guess if $\prec_{k,f}(\alpha_1,...,\alpha_k)$ holds. Clearly, this algorithm runs in non-deterministic polynomial time. $\qed$\\

We are now in the position to define the $T$-hierarchy:

\begin{definition} \em
Let $k \geq 2$ and $\Sigma$ be a set of constraints. Then $\Sigma \in T[k]$
iff there is $k' \in [k] \backslash \setone{1}$ such that for every
$\Sigma' \in part'(\Sigma,k')$ it holds that $\Sigma'$ is safe.$\punto$
\end{definition}

We call $T[k]$ the $k$-th level of the $T$-hierarchy. As a corollary from
Proposition~\ref{krest} we obtain that we can decide whether a set of constraints
is in $T[k]$ by a $\mbox{coNP}$-algorithm. We next give an example for constraints
in the $T$-hierarchy.

\begin{example} \em \label{tkecht}
We set $\Sigma_{k+1} := \setone{\alpha_{k+1}}$, where
$\alpha_{k+1} := S(x_{k+1}), R_k(x_1,...,x_{k+1}) \rightarrow \exists y R_k(y,x_1,...,x_{k})$.
It holds that $\prec_{k,\emptyset}(\alpha,...,\alpha)$ but not
$\prec_{k+1,\emptyset}(\alpha,...,\alpha)$. So the minimal $(k+1)$-restriction
system does not contain any cycle, but the minimal $k$-restriction system does.
Therefore $\Sigma_{k+1} \in T[k+1]$. On the other hand, we observe that the
constraint is not safe, so it is not contained in $T[k]$. Also note
that the constraint in Figure~\ref{fig:schemaandconstraints1} exactly corresponds
to $\Sigma_{2}$, so it is contained in level $T[3]$.$\punto$
\end{example}

The following proposition relates the levels of the $T$-hierarchy to each other
and inductive restriction.

\begin{proposition} \em Let $k \geq 2$.
\squishlist
\item $\Sigma$ is inductively restricted iff $\Sigma \in T[2]$ 
\item  $T[k] \subseteq T[k+1]$.
\item There is some $\Sigma$ such that $\Sigma \in T[k+1] \backslash T[k]$. $\punto$
\squishend
\label{prop:tkeig}
\end{proposition}

{\bf Proof Sketch.}
(a) To prove bullet one, note that both definitions coincide exactly. 
(b) Bullet two follows by definition.
(c) For bullet three we refer back to Example~\ref{tkecht}. $\qed$\\

The next result is our main contribution concerning data-independent
chase termination. It states that, for a fixed value of~$k$, 
membership in $T[k]$ guarantees polynomial time data complexity
for the chase. Again, the technical proof can be found in~\cite{msl3_2009}.

\begin{theorem}  \em \label{main-tk} \label{MAIN-TK}
Let $k \geq 2$ and $\Sigma \in T[k]$ be a fixed set of constraints. Then, there exists a
polynomial $Q \in \mathbb{N}[X]$ such that for any database instance $I$, the length
of every chase sequence is bounded by $Q(|dom(I)|)$. $\punto$
\end{theorem}

{\bf Proof Sketch:}
Let $\Sigma$ be the set of constraints under consideration. Let $C_1,...,C_n$ be the strongly connected components of the minimal $k$-restriction system of $\Sigma$. We will show the following lemma.

\begin{lemma} \em
If the chase terminates data-independently for every strongly connected component of the minimal $k$-restriction system of $\Sigma$, then it terminates data-independently for $\Sigma$. $\punto$
\end{lemma}

{\bf Proof Sketch.}
Assume that we have a database instance $I_0$ such that the chase does not terminate. We will construct an infinite chase sequence that uses only constraints from some of the $C_i$. 

We have an infinite chase sequence $S = I_0 \stackrel{\alpha_1, \overline{a}_1}{\longrightarrow} I_1 \stackrel{\alpha_2, \overline{a}_2}{\longrightarrow} \ldots$.
Without loss of generality, we can assume that 

\begin{enumerate}
	\item every constraint from $\Sigma$ fires infinitely often,
	\item that for every $j \in \mathbb{N}$ there is some $i > j$ such that $I'_{i-1} \models \alpha_{i}(\overline{a}_{i})$, where $I'_0:=I_0$, $I_{l-1} \stackrel{\alpha_l, \overline{a}_l}{\rightarrow} J_l$ for $l \neq j$ and $J_j := J_{j-1}$ and
	\item for all $i \in \mathbb{N}$ we have that $dom(I_i) \backslash dom(I_{i-1}) \cap \Delta = \emptyset$.
\end{enumerate}

This infinite chase sequence will serve as a witness for the fact that some strongly connected component of the minimal $k$-restriction system has already an infinite chase sequence. 

For every $i \in \mathbb{N} \backslash [k]$ we do the following. Let $n$ be a null value of level $i$ such that $n \in dom(I_h) \backslash dom(I_{h-1})$. So $n$ was introduced in chase step $h$ which must be due to an application of a TGD. W.l.o.g.~there is a chase step $h' \geq h$ minimal such that $\alpha_{h'}(\overline{a}_{h'})$ is violated, where $n \in \overline{a}_{h'}$ and $n$ appears in $head(\alpha_{h'})(\overline{a}_{h'})$. Then we find $\beta_1,..., \beta_{k-2}$ such that $\prec_{k,\emptyset}(\beta_1,...,\beta_{k-2},\alpha_h,\alpha_{h'})$. W.l.o.g.~there is a chase step $h'' \geq h'$ minimal such that $\alpha_{h''}(\overline{a}_{h''})$ is violated, where $n \in \overline{a}_{h''}$ and $n$ appears in $head(\alpha_{h''})(\overline{a}_{h''})$. Then we find $\beta_1,..., \beta_{k-3}$ such that $\prec_{k,f}(\beta_1,...,\beta_{k-3},\alpha_h,\alpha_{h'},\alpha_{h''})$. Iterating this procedure, we obtain a subset of $\Sigma$'s minimal $k$-restriction system.  Considering its strongly connected components, we observe that every such component is contained in some $C_i$ due to monotonicity of the construction of a minimal restriction system. Thus, there must be some $i_0 \in [n]$ for which we have an infinite chase sequence starting with the instance $I_0$. $\qed$

The theorem follows from a repeated application of the lemma. The polynomial time data complexity follows from the fact that $k$-restriction systems bound the depth of null values (see \cite{gn2008}) data-independently.
$\qed$

\subsection{An Algorithmic Approach}
\label{sec:puttingtogether}

\begin{figure}[t]
\centering
\begin{boxedminipage}{6.5cm}
\begin{tabbing}
lll \= lllll \= lllllll \= lllllllll \= lllllllllll \= \kill
{\bf sub}($\Sigma$: Set of TDGs and EGDs, $k$: not equal to $1$) \{\\
\>1:\>{\bf if}\ \ ($\Sigma$ is safe)\ \ {\bf then}\ \ \\
\>2:\>\>return true;\\
\>3:\> {\bf endif}\\
\>4:\>compute the strongly connected components (as\\
\>\> sets of constraints) $C_1$, $\dots$, $C_n$ of the minimal\\
\>\> $k$-restriction system of $\Sigma$;\\
\>5:\>{\bf if}\ \  (n == 0)\ \  {\bf then}\ \ \\
\>6:\>\> return true;\ \ \\
\>7:\> {\bf endif}\\
\>8:\>{\bf if}\ \  (n == 1)\ \  {\bf then}\ \ \\
\>9:\>\> {\bf if}\ \ ($C_1 \not = \Sigma$)\ \ {\bf then}\ \ \\
\>10:\>\>\> return {\bf check}($C_1$,$k$);\\
\>11:\>\> {\bf endif}\\
\>12:\>\> return false;\\
\>13:\> {\bf endif}\\
\>14:\>{\bf for}\ \  i=1 to n\ \  {\bf do}\ \ \\
\>15:\>\> {\bf if}\ \ ({\bf not check}($C_i$,$k$))\ \ {\bf then}\ \ \\
\>16:\>\>\> return false;\\
\>17:\>\> {\bf endif}\\
\>18:\>{\bf endfor}\\
\>19:\>return true; \}
\end{tabbing}

%\vspace{0.5cm}

\begin{tabbing}
lll \= lllll \= lllllll \= lllllllll \= lllllllllll \= \kill
{\bf check}($\Sigma$: Set of TDGs and EGDs, $k$: not equal to $1$) \{\\
\>1:\> {\bf for} $i=k$ {\bf downto} $2$ {\bf do}\\
\>2:\>\> {\bf if} $(sub(\Sigma,i))$ {\bf then} return true;\\
\>3:\> {\bf endfor}\\
\>4:\> return false; \}
\end{tabbing}
\end{boxedminipage}

%\vspace{-0.3cm}
\caption{Algorithm to decide membership in $T[\cdot]$.}
%\vspace{-0.4cm}
\label{fig:algoeff}
\end{figure}

This section aims to develop an efficient algorithm to test membership in $T[k]$.
We have seen before that the computation of $k$-restriction systems is costly
because we need \textsc{NP} time to compute the relation $\prec_{k,P}$.
For this reason, we present an algorithm that avoids the computation of $k$-restriction
systems where possible. It relies on the idea that (the weaker condition) safety
can be checked in polynomial time (cf.~Section~\ref{subsec:safety}).
Before computing the $k$-restriction system, we always check for safety and,
whenever safety holds, we conclude that the
chase for the respective constraint set terminates and omit the $k$-restriction
system computation.

To give a simple example, consider the constraint
from Example~\ref{ex:safe2}, which has been shown to be safe, and assume 
we want to test if it falls into some (fixed) level $k$ of the
$T$-hierarchy. Computing a $k$-restriction system is superfluous,
because membership in $T[k]$ trivially follows from the satisfaction
of the safety condition. 

In general, the situation is, of course, not that simple. 
Consider for instance the constraint set $\Sigma'$ from Example~\ref{ex:ir1} extended
by $\setone{\alpha_4, \alpha_5}$,
where $\alpha_4 := E(x_1,x_2) \rightarrow T(x_1,x_2)$, $\alpha_5 := T(x_1,x_2) \rightarrow T(x_2,x_1)$,
and call the resulting constraint set $\Sigma''$. Assume we want to show that $\Sigma''$
is inductively restricted (i.e., in $T[2]$).
It follows from Example~\ref{ex:notsafestratified1} that $\Sigma''$ is not safe.
In direct correspondence to
Example~\ref{ex:ir1} it follows that the minimal $2$-restriction system for
$\Sigma''$ is G'($\Sigma''$):=($\Sigma''$,\{($\alpha_1,\alpha_2$),($\alpha_2$,$\alpha_1$),($\alpha_3$,$\alpha_1$),($\alpha_3$,$\alpha_2$),($\alpha_1$,$\alpha_4$), ($\alpha_2$,$\alpha_4$),($\alpha_4$,$\alpha_5$),($\alpha_5$,$\alpha_5$)\}), where f($\alpha_1$) = f($\alpha_2$) := \{E$^1$,E$^2$,S$^1$\},
f($\alpha_3$) := $\emptyset$, f($\alpha_4$) := \{E$^1$,E$^2$\}
and f($\alpha_5$) := \{T$^1$,T$^2$\}. This $2$-restriction system contains the strongly
connected components \{$\alpha_1$,$\alpha_2$\} and \{$\alpha_5$\}.
For \{$\alpha_1$,$\alpha_2$\} we must compute its minimal $2$-restriction system because
it is not safe, but for \{$\alpha_5$\} we can avoid this complexity because
we know that $\alpha_5$ is safe (indeed it is a full TGD) and therefore the chase
terminates. We implement the scheme described above in algorithm $\textit{check}$,
provided in Figure~\ref{fig:algoeff}.

\begin{proposition} \em
Algorithm $check$ terminates and correctly decides membership in the $T$-hierarchy,
i.e.~$check(\Sigma,k)$ returns true if and only if $\Sigma \in T[k]$. $\punto$
\end{proposition}

{\bf Proof Sketch.}
The algorithm terminates because all recursive calls are made on constraint sets with size smaller than the input constraint set. What the algorithm does is trying to avoid the computation of $k$-restriction systems by testing for safeness. The correctness follows from the proof of Theorem \ref{main-tk} because the only property we need to show is that for all $\Sigma' \in part(\Sigma,k)$ the chase terminates, which is ensured by the additional safety checks.
$\qed$\\

\section{Data-dependent Termination}
\label{sec:dep}

\begin{figure}[t]
\centering
\begin{boxedminipage}{6.8cm}
\begin{tabbing}
x \= xxxxxxxxxxxxxxxx \= \kill
\>\underline{\bf Sample Schema:}\>$\textit{hasAirport}(\textit{c\_id})$\\
\>\>$\textit{fly}(\textit{c\_id1},\textit{c\_id2},\textit{dist})$\\
\>\>$\textit{rail}(\textit{c\_id1},\textit{c\_id2},\textit{dist})$\\
\\[-0.2cm]
\>\underline{\bf Constraint Set:}\>$\Sigma = \setone{\alpha_1,\alpha_2,\alpha_3}$, where
\end{tabbing}
\vspace{-0.3cm}
\begin{tabbing}
x \= xxx \= \kill
\>$\alpha_1:$\>If there is a flight connection between two cities,\ \ \ \\
\>\>both of them  have an airport:\\
\>\>$\textit{fly}(c_1,c_2,d) \rightarrow \textit{hasAirport}(c_1), \textit{hasAirport}(c_2)$\\
\\[-0.2cm]
\>$\alpha_2:$\>Rail-connections are symmetrical:\\
\>\>$\textit{rail}(c_1,c_2,d) \rightarrow \textit{rail}(c_2,c_1,d)$\\
\\[-0.2cm]
\>$\alpha_3:$\>Each city that is reachable via plane has at\\
\>\>least one outgoing flight scheduled:\\
\>\>$\textit{fly}(c_1,c_2,d) \rightarrow \exists c_3, d' \textit{fly}(c_2,c_3,d')$\\[-0.3cm]
\end{tabbing}
\end{boxedminipage}
%\vspace{-0.3cm}
\caption{Sample database schema and constraints.}
\label{fig:schemaandconstraints}
%\vspace{-0.4cm}
\end{figure}

So far, we discussed conditions that guarantee
chase termination for every database instance. In this section, we study
the problem of data-dependent termination, i.e.~given
a constraint set $\Sigma$ and a {\it fixed} instance $I$, does
the chase with $\Sigma$ terminate on $I$? By the best of our knowledge,
this problem has not been studied before. Therefore, we start our discussion
with a motivating scenario. Let us consider the travel agency
database in Figure~\ref{fig:schemaandconstraints}, where predicate \textit{hasAirport}
contains cities that have an airport and \textit{fly} (\textit{rail}) stores
flight (rail) connections between cities, including their distance \textit{dist}.
In addition to the schema, constraints $\alpha_1$-$\alpha_3$ have been specified.
For instance, $\alpha_3$ might have been added to
assert that, for each city reachable via plane, the schedule is integrated in
the local database. Now consider the CQ $q_1$ below.
% (in datalog notation, with constant $c_1$ and variables $x_1$, $x_2$, $y_1$, $y_2$).

%\vspace{-0.05cm}
\begin{tabbing}
x \= xxx \= \kill
\>$q_1$:\>$\textit{rf}(x_2) \leftarrow \textit{rail}(c_1,x_1,y_1), \textit{fly}(x_1,x_2,y_2)$
\end{tabbing}
%\vspace{-0.05cm}

The query selects all cities that can be reached from $c_1$ through
rail-and-fly. Assume that, in the style of semantic query optimization, we want to
optimize $q_1$ under constraints $\Sigma$ using the chase. We then interpret the body of $q_1$ as
database instance $I := \{\textit{rail}(c_1,x_1,y_1),\textit{fly}(x_1,x_2,y_2)\}$,
where  $c_1$ is a constant and the $x_i$, $y_i$ labeled nulls. We observe that
$\alpha_3$ does not hold on~$I$, since there is a flight to
city $x_2$, but no outgoing flight from $x_2$. Hence, the chase adds a new tuple
$t_1 := \textit{fly}(x_2,x_3,y_3)$ to $I$, where $x_3$, $y_3$ are fresh labeled
null values. In the resulting instance $I' := I \cup \{ t_1 \}$, $\alpha_3$ is
again violated (this time for $x_3$) and in subsequent steps the chase adds
$\textit{fly}(x_3,x_4,y_4)$, $\textit{fly}(x_4,x_5,y_5)$,
$\textit{fly}(x_5,x_6,y_6)$, $\dots$. Obviously, it will never terminate.

Arguably, reasonable applications should never risk non-termination. 
It is clear, though, that the existence of (even a single)
non-terminating chase sequences also means that no data-independent
termination condition holds. Hence, based
on data-independent conditions no query at all
could be safely chased with the constraint set from Figure~\ref{fig:schemaandconstraints} and the benefit of the chase algorithm would be completely
lost.\footnote{Note that, principally, query optimization could also be done
with a bounded portion of the chase result, but in general we do not find minimal
rewritings of the input query in the style of~\cite{dpt2006}. Therefore,
it is desirable to guarantee chase termination.} 
Despite the fact that there is a non-terminating chase sequence, however,
there might be queries for which the
chase with the constraint set from Figure~\ref{fig:schemaandconstraints}
terminates.
Tackling such situations, we propose to investigate data-dependent
chase termination, i.e.~to study sufficient termination guarantees
for a {\it fixed instance} when no general termination guarantees apply.
We illustrate the benefits of having such guarantees for query
$q_2$ below, which selects all cities $x_2$ that can be reached from $c_1$ via
rail-and-fly and the same transport route leads back from $x_2$ to $c_1$
(where $c_1$ is a constant and the $x_i$, $y_i$ are variables).

%\vspace{-0.05cm}
\begin{tabbing}
x \= xxx \= xxxxxxxxx \= \kill
\>$q_2$:\>$\textit{rffr}(x_2) \leftarrow$\>$\textit{rail}(c_1,x_1,y_1), \textit{fly}(x_1,x_2,y_2),$\\
\>\>\>$\textit{fly}(x_2,x_1,y_2), \textit{rail}(x_1,c_1,y_1)$
\end{tabbing}
%\vspace{-0.05cm}

Query $q_2$ violates only $\alpha_1$. It is easy to verify that 
the chase terminates for this query and transforms $q_2$
into $q_2'$:

%\vspace{-0.05cm}
\begin{tabbing}
x \= xxx \= xxxxxxxxx \= \kill
\>$q_2'$:\>$\textit{rffr}(x_2) \leftarrow$\>$\textit{rail}(c_1,x_1,y_1), \textit{fly}(x_1,x_2,y_2),$\\
\>\>\>$\textit{fly}(x_2,x_1,y_2), \textit{rail}(x_1,c_1,y_1),$\\
\>\>\>$\textit{hasAirport}(x_1), \textit{hasAirport}(x_2)$
\end{tabbing}
%\vspace{-0.05cm}

The resulting query $q_2'$ satisfies all constraints and is a so-called 
{\it universal plan}~\cite{dpt2006}: intuitively, it incorporates all possible
ways to answer the query. As discussed in~\cite{dpt2006}, the universal plan
forms the basis for finding smaller equivalent queries (under the respective constraints),
by choosing any subquery of $q_2'$ and testing if it can be chased to a
homomorphical copy of $q_2'$. Using this technique we can easily
show that the following two queries are equivalent to $q_2$.

%\vspace{-0.05cm}
\begin{tabbing}
x \= xxx \= xxxxxxxxx \= \kill
\>$q_2''$:\>$\textit{rffr}(x_2) \leftarrow$\>$\textit{rail}(c_1,x_1,y_1), \textit{fly}(x_1,x_2,y_2),$\\
\>\>\>$\textit{fly}(x_2,x_1,y_2)$\\
\>$q_2'''$:\>$\textit{rffr}(x_2) \leftarrow$\>$\textit{hasAirport}(x_1),\textit{rail}(c_1,x_1,y_1),$\\
\>\>\>$\textit{fly}(x_1,x_2,y_2), \textit{fly}(x_2,x_1,y_2)$
\end{tabbing}
%\vspace{-0.05cm}

Instead of $q_2$ we thus could evaluate $q_2''$ or $q_2'''$,
which might well be more performant: in both $q_2''$ and $q_2'''$
the join with $\textit{rail}(x_1,c_1,y_1)$ has been eliminated; moreover,
if $\textit{hasAirport}$ is duplicate-free,  the additional join
of $\textit{rail}$ with $\textit{hasAirport}$ in $q_2'''$ may serve as a filter that
decreases the size of intermediate results and speeds up query evaluation.
This strategy is called {\it join introduction} in SQO (cf.~\cite{k1986}).
Ultimately, the chase for $q_2$ made it possible to detect 
$q_2''$ and $q_2'''$, so it would be desirable to have data-dependent termination
guarantees that allow us to chase~$q_2$ (and $q_2''$, $q_2'''$). We will present
such conditions in the remainder of this section.

\subsection{Static Termination Guarantees} 

Our first approach to data-dependent chase termination is a static one. It
relies on the observation that the chase will always terminate on instance $I$
if the subset of constraints that might fire when chasing $I$ with $\Sigma$
is contained in some level of the $T$-hierarchy. We call a constraint
$\alpha \in \Sigma$ {\it $(I,\Sigma)$-irrelevant} if and only if
there is no chase sequence such that $\alpha$ can eventually fire,
i.e.~no chase sequence of the form
$I \stackrel{\alpha_1, \overline{a_1}}{\longrightarrow} \dots \stackrel{\alpha,\overline{a}}{\longrightarrow} \dots$.
%and formalize our observation in Lemma~\ref{lem:relterm}.

\begin{lemma}\em \label{LEM:RELTERM} \label{lem:relterm}
Let $k \geq 2$ and $\Sigma' \subseteq \Sigma$ s.t. $\Sigma \setminus \Sigma'$ is a set of $(I,\Sigma)$-irrelevant constraints. If $\Sigma' \in T[k]$, then the chase with
$\Sigma$ terminates for instance $I$.$\punto$
\end{lemma}

{\bf Proof Sketch.}
It holds that $\Sigma'$ contains all constraints that may fire during the execution of the chase starting with $I$ and $\Sigma$. $I^{\Sigma'}$ is finite and $I^{\Sigma'}=I^{\Sigma}$.$\qed$\\

Hence, the crucial point is to effectively compute the set of $(I,\Sigma)$-irrelevant
constraints. Unfortunately, it turns out that checking $(I,\Sigma)$-irrelevance is
an undecidable problem in general:

\begin{theorem} \em
Let $\Sigma$ be a set of constraints, $\alpha \in \Sigma$ a constraint,
and $I$ an instance. It is undecidable if $\alpha$ is $(I,\Sigma)$-irrelevant.$\punto$
\label{th:irr} \label{TH:IRR}
\end{theorem}

{\bf Proof Sketch.}
It is well-known that the following problem is undecidable: given a Turing machine $M$ and and a state transition $t$ from the description of $M$, does $M$ reach $t$ (given the empty string as input)? From $(M,t)$, we will compute a set of TGDs and EGDs $\Sigma_M$ and a TGD $\alpha_t \in \Sigma_M$ such that the following equivalence holds: $M$ reaches $t$ (given the empty string as input) $\Leftrightarrow$ there is a chase sequence in the computation of the chase with $\Sigma_M$ applied to the empty instance such that $\alpha_t$ will eventually fire.

Our reduction uses the construction in the proof of Theorem~1 in~\cite{dnr2008}.  To be self-contained, we review it here again. We use the signature consisting of the relation symbols: $T(x, a, y)$ tape ``horizontal'' edge from $x$ to $y$ with symbol $a$; $H(x, s, y)$ head ``horizontal'' edge from $x$ to $y$ with state $s$; $L(x, y)$ left ``vertical'' edge; $R(x, y)$ right ``vertical'' edge; 
$A_{\delta}(x),B_{\delta}(x)$ for every stater transition $\delta$, one constant for every tape symbol, one constant for every head state, the special constant $B$ marking the beginning of the tape and $\square$ to denote an empty tape cell. The set of constraints $\Sigma_M$ is as follows.

\begin{enumerate}
	\item The initial configuration:\\ $\exists w, x, y, z T(w,B, x), T (x,\square, y),H(x, s_0, y),T(y,E,z)$\\
where $\square$ is the blank symbol and $s_0$ is the initial state (both are constants).

\item For every state transition $\delta$ which moves the head to the right,
replacing symbol $a$ with $a'$ and going from state $s$ to state $s'$:\\
$T(x, a, y),H(x, s, y), T (y, b, z) \rightarrow$\\
$\exists x', y', z' L(x, x'), R(y, y'), R(z, z'), T(x', a', y'),$\\
 $T(y', b, z'),H(y', s', z'),A_{\delta}(w')$.\\
 Here $a, s, a', b$, and $s'$ are constants.

\item For every state transition $\delta$ which moves the head to the right
past the end of the tape replacing symbol a with a' and going
from state s to state s':\\
$T(x, a, y),H(x, s, y), T (y, E, z) \rightarrow$\\
$\exists w', x', y', z' L(x, x'), R(y, y'), R(z, z'), T(x', a', y'),$\\
 $T(y', \square, z'),H(y', s', z'), T(y',E,w'),A_{\delta}(w')$.\\
 Here $a, s, a', b$, and $s'$ are constants.

\item Similarly for state transitions which move the head to the left.

%$T(x, b, y), T (y, a, z), H(y, s, z) \rightarrow$\\
%$\exists x', y', z' L(x, x'), L(y, y'), R(z, z'), T(x', b, y'), H(x', s', y'), T(y', a', z')$.\\
% Here $a, s, a', b$, and $s'$ are constants.

\item Similarly for state transitions which do not move the head.

\item For every state transition $\delta$:\\
$A_{\delta}(x) \rightarrow B_{\delta}(x)$

%\item Create empty tape cells if necessary, while ensuring consistency with respect to existing tape cells:\\
%$T(x,z,y) \rightarrow \exists z T(y,\square,z)$.\\
%$T(x,z,y), T(x,z',y) \rightarrow z = z'$.\\
%Here $z, z'$ are  variables.

\item Left copy:\\
$T(x, a, y), L(y, y') \rightarrow \exists x' L(x, x'), T(x', a, y')$.\\
 Here $a$ is a  constant.

\item Right copy:\\
$T(x, a, y),R(x, x') \rightarrow \exists y' T(x', a, y'),R(y, y')$.\\
 Here $a$ is a  constant.
\end{enumerate}

The state transition $t$ is transformed to $\alpha_t$ in the same way like in bullet six above. It is crucial to the proof that every state transition $\delta$ in $M$ is represented as a single TGD $A_{\delta}(x) \rightarrow B_{\delta}(x)$. The constraint for the initial configuration fires exactly once. The computation of the chase with this set of constraint can be understood as a grid and each row in the grid represents a configuration of the Turing machine. It can be shown that $(M,t)$ is a yes-instance if and only if $(\Sigma_M,\alpha_t)$ is a yes-instance. Thus, the equivalence from above holds. $\qed$\\

This result prevents us from computing the minimal set of constraints that may
fire when chasing $I$. Still, we can give sufficient conditions that guarantee
$(I,\Sigma)$-irrelevance for a constraint. For this purpose, we use the chase graph. 
%The chase graph for $\Sigma$ is
%a graph $G(\Sigma) =(\Sigma,\prec)$, where $\alpha \prec \beta$ holds for
%$\alpha, \beta \in \Sigma$ iff the first three bullets from Definition \ref{def-verf}
%hold. It was shown in \cite{dnr2008} that, given $\Sigma$, the chase graph can be
%computed by an $\textsc{NP}$-algorithm.

\begin{proposition} \label{prop:bla} \label{PROP:BLA} \em Let $I$ be an instance and $\Sigma$ be a set of constraints such that every constraint has a non-empty body. Further
let $\alpha_I := \exists \overline{x} \bigwedge_{R(\overline{x}') \in I}  R(\overline{x}')$
where $\overline{x} := \bigcup_{R(\overline{x}') \in I} \overline{x}'$.
If the c-chase graph $G_c(\Sigma \cup \setone{\alpha_I})$ contains no directed path from 
 $\alpha_I$ to $\beta \in \Sigma$, then $\beta$ is $(I,\Sigma)$-irrelevant.$\punto$
\end{proposition}

{\bf Proof Sketch}
Assume that $\beta$ is not $(I,\Sigma)$-irrelevant. Then, there is a chase sequence $I \stackrel{\alpha_1, \overline{a_1}}{\longrightarrow} I_1 \stackrel{\alpha_2, \overline{a_2}}{\longrightarrow} \dots \stackrel{\alpha_r, \overline{a_r}}{\longrightarrow} I_r \stackrel{\beta,\overline{a}}{\longrightarrow} \dots$.
If $\alpha_I \prec_c \beta$ we are finished. Otherwise, there must be some $n_r \in [r]$ such that $\alpha_{n_r} \prec_c \beta$ (otherwise $\beta$ could not fire). If $\alpha_I \prec_c \alpha_{n_r}$ we are finished. Otherwise, there must be some $n_{r-1} \in [n_r - 1]$ such that $\alpha_{n_{r-1}} \prec_c \alpha_{n_r}$ (otherwise $\alpha_{n_r}$ could not fire). After some finite amount of iterations of this process we have that $\alpha_I \prec_c \alpha_{n_1} \prec_c ... \prec_c \alpha_{n_r} \prec_c \beta$. Therefore, the chase graph contains a directed path from $\alpha_I$ to $\beta$. $\qed$\\

Proposition~\ref{prop:bla} together with Lemma~\ref{lem:relterm} gives us a
sufficient data-dependent condition for chase termination, as illustrated in the
following example.
 
\begin{example} \em
Consider constraint set $\Sigma$ from Figure~\ref{fig:schemaandconstraints} and 
$q_2$ from the beginning of this section. We set 

\begin{tabbing}
xx \= xxxxxxxxxxxxxxxxxxxxxxx \= \kill
\>$\alpha_I := \exists c_1,x_1,x_2,y_1,y_2$\>$\textit{rail}(c_1,x_1,y_1), \textit{fly}(x_1,x_2,y_2),$\\
\>\>$\textit{fly}(x_2,x_1,y_2), \textit{rail}(x_1,c_1,y_1)$
\end{tabbing}

and compute the chase graph

\begin{tabbing}
xx \= \kill
\>$G(\Sigma \cup \setone{\alpha_I}) := (\Sigma \cup \setone{\alpha_I}, \{(\alpha_I,\alpha_1), (\alpha_3,\alpha_3)\})$.
\end{tabbing}

By Proposition~\ref{prop:bla}, $\alpha_2$ and $\alpha_3$ are $(I,\Sigma)$-irrelevant. It holds
that $\Sigma \setminus \setone{\alpha_2,\alpha_3} = \setone{\alpha_1}$
is inductively restricted, so we know from Lemma~\ref{lem:relterm} that the
chase of $q_2$ with $\Sigma$ terminates. Similar argumentations hold for
$q_2''$ and $q_2'''$ from the beginning of Section~\ref{sec:dep}.$\punto$
\end{example}
 
\subsection{Monitoring Chase Execution} 

If the previous data-dependent termination condition does not apply, we propose
to monitor the chase run and abort if tuples are created that may potentially
lead to non-termination, an approach that is dynamical by nature.
We introduce a data structure called {\it monitor graph}, which
allows us to track the chase run.

\begin{definition} \em
A {\it monitor graph} is a tuple $(V,E)$, where $V \subseteq \Delta_{null} \times 2^{\mbox{pos}(\Sigma)}$ and $E \subseteq V \times \Sigma \times 2^{\mbox{pos}(\Sigma)} \times V$. $\punto$
\end{definition}

A node in a monitor graph is a tuple $(n,\pi)$, where $n$ is a null value and
$\pi$ the set of positions in which $n$ was first created (e.g. as null value
with the help of some TGD). An edge $(n_1,\pi_1,\varphi_i,\Pi,n_2,\pi_2)$ between
$(n_1,\pi_1)$, $(n_2,\pi_2)$ is labeled with the constraint $\varphi_i$
that created $n_2$ and the set of positions $\Pi$ from the body of $\varphi_i$
in which $n_1$ occurred when $n_2$ was created. The monitor graph is successively
constructed while running the chase, according to the following definition.

\begin{definition} \em
The monitor graph $G_{\mathcal{S}} := G_{r}$ w.r.t. 
$\mathcal{S} = I_0 \stackrel{\varphi_0, \overline{a}_0}{\longrightarrow} 
\ldots \stackrel{\varphi_{r-1}, \overline{a}_{r-1}}{\longrightarrow} 
I_r$ is a monitor graph that is inductively defined as follows
\squishlist
	\item $G_0 = (\emptyset,\emptyset)$ is the empty chase segment graph.
	\item If $i < r$ and $\varphi_i$ is an EGD then $G_{i+1} := G_i$.
	\item If $i < r$ and $\varphi_i$ is a TGD then $G_{i+1}$ is obtained from $G_i = (E_i,V_i)$ as follows. If the chase step $I_i \stackrel{\varphi_i, \overline{a}_i}{\longrightarrow} I_{i+1}$  does not introduce any new null values, then $G_{i+1} := G_{i}$. Otherwise, $E_{i+1}$ is set as the union of $E_i$ and all pairs $(n,\pi)$, where $n$ is a newly introduced null value and $\pi$ the set of positions in which $n$ occurs. $V_{i+1} := V_i \cup \set{(n_1,\pi_1,\varphi_i,\Pi,n_2,\pi_2)}{(n_1,\pi_1) \in E_i, (n_2,\pi_2) \in E_{i+1} \backslash E_i \text{ and } \Pi \text{ is the set of positions in } body(\varphi_i(\overline{a}_i))\ \ \text{where } \\ n_1 \text{ occurs}}$.  $\punto$
\squishend
\end{definition}

The size of the monitor graph is polynomial in the length
of the chase sequence plus the length of the constraints' encoding. We illustrate
the definition of the chase graph by a small example.

 \begin{example} \em
 Consider the constraint $\Sigma_3 = \setone{\alpha_3}$, where $\alpha_3 := S(x_{3}), R_k(x_1,x_2,x_{3}) \rightarrow \exists y R_k(y,x_1,x_{2})$ from Example~\ref{tkecht}. Assume we have an instance of the form $I_0 := \setone{S(a_1), S(a_2), S(a_3), E(a_1,a_2,a_3)}$. Then, the only chase sequence is $I_0 \rightarrow I_1 \rightarrow I_2 \rightarrow I_3$, where 
 $I_1 = I_0 \cup \setone{E(y_1,a_1,a_2)}$, 
 $I_2 = I_1 \cup \setone{E(y_2,y_1,a_1)}$
 $I_3 = I_2 \cup \setone{E(y_3,y_2,y_1}$. As $y_1$ is not in relation $S$ the chase terminates.
 The monitor graph contains the path $(y_1,\setone{E^1}) \stackrel{\alpha_3,E^1}{\longrightarrow} (y_2,\setone{E^1}) \stackrel{\alpha_3,E^1}{\longrightarrow} (y_3,\setone{E^1})$ plus
an additional edge $(y_1,\setone{E^1}) \stackrel{\alpha_3,E^2}{\longrightarrow} (y_3,\setone{E^1})$. $\punto$ \label{ex:cyclic}
 \end{example}
 
Our next task is to define a necessary criterion for non-termination on top of the
monitor graph. To this end, we introduce the notion of {\it k-cyclicity}.

\begin{definition} \em
Let $G = (V,E)$ be a monitor graph and $k \in \mathbb{N}$. $G$ is called $k$-cyclic if and only if there are pairwise distinct edges $v_1,...,v_k \in E$ such that 
\squishlist
\item there is a path in $E$ that sequentially contains $v_1$ to $v_k$ and
\item for all $i \in [k-1]$: $p_{2,3,4,6}(v_i) = p_{2,3,4,6}(v_{i+1})$.$\punto$
\squishend
\end{definition}

\begin{example} \em
Consider the scenario from Example \ref{ex:cyclic}.
According to the previous definition, the chase graph presented there is
$2$-cyclic, but not $3$-cyclic. $\punto$
 \end{example}

We call a chase sequence {\it $k$-cyclic} if its monitor graph is $k$-cyclic.
A chase sequence may potentially be infinite if some finite prefix is $k$-cyclic,
for any $k \geq 1$: 

\begin{lemma} \label{dynamic} \label{DYNAMIC} \em
Let $k \in \mathbb{N}$. If there is some infinite chase sequence $\mathcal{S}$
when chasing $I_0$ with $\Sigma$, then there is some finite prefix of
$\mathcal{S}$ that is $k$-cyclic. $\punto$
\end{lemma}

{\bf Proof:} Assume that 
\begin{itemize}
 \item we have an infinite chase sequence $\mathcal{S} = (I_i)_{i \in \mathbb{N}}$ and 
 \item there is some $k \in \mathbb{N}$ such that every finite prefix of $\mathcal{S}$ is not $k$-cyclic.
\end{itemize}

Let $(\mathcal{S}_i)_{i \in \mathbb{N}}$ be the sequence of finite prefixes of $\mathcal{S}$ (such that $\mathcal{S}_i$ is a chase sequence of length $i$) and let $(G_{\mathcal{S}_i})_{i \in \mathbb{N}}$ the respective sequence of monitor graphs. A path in a monitor graph is a finite sequence of edges $e_1,...,e_l$ (and not of nodes) such that $p_{5,6}(e_i)=p_{1,2}(e_{i+1})$ for $i \in [l-1]$. 

We define the notion of {\it depth} of a node in a monitor graph. Let $v$ be a node in $G_{\mathcal{S}_i}$ and $pred(v)$ the set of predecessors of $v$. In case $v$ has no predecessors, the depth of $v$, $depth_{G_{\mathcal{S}_i}}(v)$, is defined as zero. In case $v$ has predecessors, then $depth_{G_{\mathcal{S}_i}}(v) := 1 + max \set{depth_{G_{\mathcal{S}_i}}(w)}{w \in pred(v)}$.\\

The following claim follows immediately from the definition of the monitor
graph. The formal proof is left to the reader.

\begin{proposition}\em \label{basic}
Let $v$ be a node in $G_{\mathcal{S}_i}$ and $j > i$.

\squishlist
 \item $G_{\mathcal{S}_i}$ is an acyclic labeled tree.
 \item Every null value that appears in $I_i$ appears in some first position of a node in $G_{\mathcal{S}_i}$.
	\item There is a homomorphism\footnote{A homomorphism leaves relational symbols and constraints untouched, i.e.~is the identity on elements from $\Delta$.} $h_{ij}$ from $G_{\mathcal{S}_i}$ to $G_{\mathcal{S}_j}$ such that $depth_{G_{\mathcal{S}_i}}(v) \leq depth_{G_{\mathcal{S}_j}}(h_{ij}(v))$.
	\item If $I_i \stackrel{\varphi_i, \overline{a}_i}{\rightarrow}I_{i+1}$, $b \in \overline{a}_i$ is a null value and $c$ is a null value that was newly created in this step, then the depth of any node in $G_{\mathcal{S}_{i+1}}$ in which $b$ appears is strictly smaller than the depth of any node in $G_{\mathcal{S}_{i+1}}$ in which $c$ appears. (The proof is by induction on $i$.) $\punto$
\squishend
\end{proposition}
	 
	The next proposition is the most important step in the proof of this lemma and follows directly from bullet four in Proposition~\ref{basic}.
	 
	\begin{proposition}\em \label{adv} Let $i \in \mathbb{N}$. For every  $d \in \mathbb{N} \cup \setone{0}$ there is a number $k_d \in \mathbb{N}$ such that for every $i \in \mathbb{N}$ it holds that $|\set{v}{depth_{G_{\mathcal{S}_i}}(v) \leq d}| \leq k_d$. \textit{Note that $k_d$ is independent from $i$.} (The proof is by induction on $d$.) $\punto$
	 \end{proposition}

We observe another fact.
	\begin{proposition}\em
There is some $p_k \in \mathbb{N}$ such that if some $G_{\mathcal{S}_i}$ has a path of length $p_k$, then $\mathcal{S}_i$ is $k$-cyclic. $\punto$
\end{proposition}

This is because we have only a bounded number of relational symbols and constraints available. The remaining step in the proof is to show that if we choose~$i$ large enough, then $G_{\mathcal{S}_{i}}$ contains a path of length $p_k$. Assume that this claim does not hold. By Proposition~\ref{adv}, the number of nodes of a certain depth is bounded (independent of $i$). So, if for any $i$ there would be no path of length $p_k$ in $G_{\mathcal{S}_{i}}$, then the number of nodes in $G_{\mathcal{S}_{i}}$ would be bounded (independent of $i$). This implies that the chase has introduced only a bounded number of fresh null values, which contradicts to the assumption of an infinite chase sequence. $\qed$

To avoid non-termination, an application can fix a cycle-depth $k$ and stop the chase when
this limit is exceeded. For every terminating chase sequence there is a
$k$ such that the sequence is not $k$-cyclic, so if $k$ is chosen large enough the chase
will succeed. We argue that $k$-cyclicity is a {\it natural} condition that considers
situations that may cause non-termination, so this approach is preferable to
blindly chasing the instance and stopping after a fixed amount of chase steps. As
justified by the following proposition, applications can choose $k$ following a
pay-as-you-go principle: for larger $k$-values the chase succeeds in more cases.

\begin{proposition} \em \label{prop:nutzendynamic} \label{PROP:NUTZENDYNAMIC}
For $k \in \mathbb{N}$ there is  $\Sigma_k$, $I_k$ such that (a)~both $\Sigma_k$ and
the subset of constraints in $\Sigma_k$ that are not $(I_k,\Sigma_k)$-irrelevant are not inductively restricted; (b) every chase sequence for $I_k$ with $\Sigma_k$ is $(k-1)$-, but not $k$-cyclic.$\punto$
 \end{proposition}

{\bf Proof:} We set $\Sigma_k := \setone{\varphi}$ and $I_k = \setone{S(c_1),...,S(c_k), R_k(c_1,...,c_k)}$, where $\varphi := S(x_k), R_k(x_1,...,x_k) \rightarrow \exists y R_k(y,x_1,...,x_{k-1})$.

First observe that $\Sigma_k$ contains no $(I,\Sigma_k)$-irrelevant constraints, so the subset of the constraints in $\Sigma_k$ that is not $(I,\Sigma)$-irrelevant equals to $\Sigma_k$. It is easy to verify that $\Sigma_k$ is not inductively restricted, although the chase with $\Sigma_k$ always terminates, independent of the underlying data instance, so condition~(a) holds. 
 
We now chase $I_k$ with $\Sigma_k$. There is only one possible chase sequence
$(J_i)_{0 \leq i \leq k}$, defined as $J_0 := I_k$, for
$i \leq k$: $J_{i} := J_{i-1} \cup \setone{R(n_i,...,n_1,c_1,...,c_{k-i})}$,
 and $n_1,...,n_k$ are fresh null values. It holds that $J_k \models \Sigma_k$.
 
 The monitor graph w.r.t.~$(J_i)_{0 \leq i \leq k}$ is $(V,E)$, where

\begin{tabbing}
xx \= \kill
\>$V:=\set{(n_i,R_k^1)}{i \in [k]}$ and\\
\>$E:=\set{(n_i,R_k^1,\varphi,R_k^{j-i},n_j,R_k^1)}{1 \leq i<j \leq k}$.
\end{tabbing}

We observe that the sequence is $(k-1)$-cyclic because $(n_1,R_k^1,\varphi,R_k^{1},n_2,R_k^1),...,(n_{k-1},R_k^1,\varphi,R_k^{1},n_k,R_k^1)$ constitute a path in the chase graph that satisfies the conditions of the definition of $(k-1)$-cyclicity. The chase sequence is not $k$-cyclic because there is no path of length at least $k$ in the monitor graph. This proves part~(b) of the proposition. $\qed$

\section{An Application}
\label{sec:apps}

Answering Conjunctive Queries on knowledge bases has recently gained
attraction~\cite{cgk2008,cgl2009}. Such knowledge bases typically have
a set of constraints associated, which imply additional tuples that are not
materialized in the knowledge base itself. An important problem is
query answering on the implied knowledge base. If the chase with these
constraints terminates, query answering can be done by answering it
on the chased knowledge. However, if no termination guarantees for the
chase can be made, more sophisticated techniques for query answering are required.
This problem was first considered in~\cite{jk1982} and then generalized
in~\cite{cgk2008} and~\cite{cgl2009}. In this section we leverage the methods
developed in Section~\ref{sec:ind}, showing that they can be used to make
the algorithms given in~\cite{cgk2008,cgl2009} applicable to broader classes
of constraints. % and not to improve the algorithms themselves. 

In \cite{cgk2008} the class of so-called weakly guarded TGDs was introduced,
which make query answering under constrained databases decidable. We first review
this notion. Later, we will generalize weakly guarded TGDs with our methods. Our starting point
is the definition of treewidth.\\

\begin{definition} \label{def:wgtgd} \em
Let $\Sigma$ be a set of TGDs.  We call $\Sigma$ {\it weakly guarded} if for every $\alpha \in \Sigma$ there exists $g_{\alpha} \in body(\alpha)$ such that for any $\pi \in aff(\Sigma) \cap pos(\alpha)$ and every variable $x_{\pi}$ that occurs in $\pi$ it holds that $x_{\pi}$ occurs also in $g_{\alpha}$. $\punto$
\end{definition}

If $\Sigma$ is weakly guarded, we abbreviate this by $\textit{WGTGD}(\Sigma)$. It was
first shown in \cite{cgk2008} that if $\textit{WGTGD}(\Sigma)$, then answering
Conjunctive Queries on $I^{\Sigma}$ is decidable for every database instance $I$,
even though $I^{\Sigma}$ may be infinite. 
Although not stated explicitly, it follows from the proof of Lemma~27 in~\cite{cgk2008}
that the crucial property for decidability of query answering of WGTGDs is that in
every chase step there is an atom in the body of the constraint under consideration that
contains all labeled nulls. We state this observation more precisely in the
following definition.

\begin{definition} \em
Let $\mathcal{S}$ be a chase sequence starting with the instance $I$. $\mathcal{S}$ has the
guarded null property if for every chase step
$I' \stackrel{\alpha, \overline{a}}{\longrightarrow} I''$ in $\mathcal{S}$ there is an
atom in $body(\alpha)(\overline{a})$ that contains every element from
$(\overline{a} \cap \Delta_{null})\backslash dom(I)$ that occurs in $head(\alpha)(\overline{a})$. $\punto$
\end{definition}

With this definition at hand we can generalize Lemma 27 in~\cite{cgk2008} to
the following version, which follows implicitly from the proof of
Lemma~27 in~\cite{cgk2008}. 

Next, we need to introduce the notion of treewidth. A hypergraph is a pair $\mathcal{H} = (V,H)$, where $H \subseteq 2^V$. The Gaifman graph of a hypergraph $\mathcal{H}$, $\mathcal{G}_{\mathcal{H}}$, has the same set of nodes like the hypergraph and contains an edge between two nodes $(v_1,v_2)$, whenever there is some $h \in H$ such that $v_1,v_2 \in h$. A tree decomposition of a graph $\mathcal{G}=(V,E)$ is a pair $(T,B)$, where $T =(N,A)$ is a graph and $B: N \rightarrow 2^V$ such that (i) $B(N) = V$, (ii) for every $(v_1,v_2) \in E$ there is some $n \in N$ such that $\setone{v_1,v_2} \in B(n)$, and (iii) for every $v \in V$ the set $\set{n \in N}{v \in B(n)}$ is the set of nodes of a connected subtree of $T$. The width of $(T,B)$ is $max \set{|B(n)|-1}{n \in N}$. The treewidth of $\mathcal{G}$, $tw(\mathcal{G})$, is the minimum width of all tree decompositions of $\mathcal{G}$. The treewidth of a hypergraph is defined as the treewidth of its Gaifman graph. Therefore the treewidth of a database instance is defined. Note that we consider an atom $R(a,b,c,d)$ as the set $\setone{a,b,c,d}$ here. 
We denote by $tw(I^{\Sigma})$ the treewidth of $I^{\Sigma}$.

\begin{lemma} $ $ \em
If all chase sequences w.r.t. $\Sigma$ and $I$ have the guarded null
property, then
$tw(I^{\Sigma}) \leq |dom(I)| + max \set{ar(R)}{R \in \mathcal{R}}$. $\punto$
\end{lemma}

Straightforwardly, we obtain the following theorem that is obtained from a result in \cite{c1989} and  the observation that in case that all chase sequences have the guarded null property, then  if $I^{\Sigma} \wedge Q$ and $I^{\Sigma} \wedge \neg Q$ are satisfiable, they have models of finite treewidth (because $I^{\Sigma}$ has such a model). 

\begin{theorem} $ $ \em \label{decgnp}
There is an algorithm that, for every set of TGDs $\Sigma$, Conjunctive Query $q$ and database instance $I$ such that every chase sequence has the guarded null property, correctly computes $q(I^{\Sigma})$. $\punto$
\end{theorem}

Unfortunately, it is not known if it is decidable if all chase sequences have
the guarded null property (given $\Sigma$ and $I$ as input), which justifies the
research regarding sufficient syntactic restrictions on the constraint set
such that all chase sequences with this constraint have the guarded null property. 
%
%\begin{theorem} \em
%Given $\Sigma,I$ as input it is undecidable if
%\squishlist
%	\item there is a chase sequence with the guarded null property.
%	\item all chase sequences have the guarded null property. $\punto$
%\squishend
%\label{th:irr2} \label{TH:IRR2}
%\end{theorem} 

The notion of affected positions is a rough syntactic overestimation on
where labeled nulls may occur in a constraint body during the execution
of the chase. With the help of $2$-restriction systems, we can improve
this overestimation. The following definition states that every TGD
$\alpha$ must have an atom in its body that contains all variables
occurring in $f(\alpha)$, where $f$ is the function from the constraint
set's minimal restriction system (cf.~Definition~\ref{rest}).
Intuitively, $f(\alpha)$ defines the set of positions in which null
values may occur during the execution of the chase.

\begin{definition} \label{def:rgtgd} \em
Let $\Sigma$ be a set of TGDs and $G'(\Sigma) = (G',f)$ its minimal
$2$-restriction system.  We call $\Sigma$ {\it restrictedly guarded}
if for every $\alpha \in \Sigma$ there exists $g_{\alpha} \in body(\alpha)$
such that for any $\pi \in f$ and every universally quantified variable $x_{\pi}$ in $body(\alpha)$
that occurs in $\pi$ it holds that $x_{\pi}$ occurs also in $g_{\alpha}$.$\punto$
\end{definition}

We call $g_{\alpha}$ a restricted guard and write $\textit{RGTGD}(\Sigma)$
to denote that constraint set $\Sigma$ is restrictedly guarded.

\begin{example} \em \label{ex:notwgtgd}
Consider the set of constraints $\Sigma := \setone{\alpha_1,\alpha_2,\alpha_3}$, where
$\alpha_1 := R(x_1,x_2), S(x_1,x_2) \rightarrow \exists y S(x_2,y)$, 
$\alpha_2 := S(x_1,x_2), S(x_3,x_1) \rightarrow R(x_2,x_1)$ and
$\alpha_3 := T(x_1,x_2) \rightarrow \exists y S(y,x_2)$. The set of affected positions
is $\textit{aff}(\Sigma)= \setone{S^1,S^2,R^1,R^2}$ and therefore $\alpha_2$
violates the condition for weak guardedness because there is no atom that
contains $x_1,x_2,x_3$. However, the constraint set is restrictedly guarded.
The minimal $2$-restriction system $((\Sigma,E),f)$ contains only the
edges $E(\alpha_1,\alpha_2), E(\alpha_3,\alpha_2)$ (and no other edges) and we have that
$f:=\setone{S^2,R^1}$.
The body of $\alpha_2$ contains the atom $S(x_1,x_2)$ which serves
as its restricted guard. $\punto$
\end{example}

Next, we relate restricted guardedness to weak guardedness and also show the crucial property that restricted guardedness ensures the guarded null property.

\begin{lemma}  \em
Let $\Sigma$ be a set of TGDs.
\squishlist
	\item $\textit{WGTGD}(\Sigma)$ implies $\textit{RGTGD}(\Sigma)$.
	
	\item There is some $\Sigma$ s.t. $\textit{RGTGD}(\Sigma)$, but not $\textit{WGTGD}(\Sigma)$.
	
	\item For every database $I$ it holds that if $\textit{RGTGD}(\Sigma)$, then every chase sequence with $\Sigma$ and $I$ has the guarded null property. $\punto$
\squishend
\label{lemma:rgwgtgd}
\end{lemma}

{\bf Proof Sketch.}
Let $(G',f)$ be the minimal $2$-restriction system for $\Sigma$. We can show by induction on the number of steps needed to compute it that $f \subseteq \textit{aff}(\Sigma)$. This implies bullet one. Bullet two is proven by Example~\ref{ex:notwgtgd}. Bullet three follows from the observation that
if a constraint $\beta$ is violated during the execution of the chase, say $J \not \models \beta(\overline{a})$, then every $(\overline{a} \cap \Delta_{null}) \backslash dom(I)$ appears in some position $g_{\beta}^i$ of some restricted guard $g_{\beta}$ in the body of $\beta$. From the construction of the minimal $2$-restriction system it follows that $g_{\beta}^i \in f$.$\qed$\\

As our final result, Lemma~\ref{lemma:rgwgtgd}  and Theorem~\ref{decgnp} imply:

\begin{corollary} \em
There is an algorithm that, for every $\textit{RGTGD}(\Sigma)$, Conjunctive Query $q$ and  database instance $I$, correctly computes $q(I^{\Sigma})$. $\punto$
\end{corollary}

\section{Conclusions}
\label{sec:conclusion}
We studied the termination of the well-known chase algorithm. By the best
of our knowledge, this was the first study that -- in addition
to the constraints -- takes the specific instance (respectively query)
into account. We also started to study the area of sufficient termination conditions for the chase which ensure, independently of the underlying data, the termination of at least one chase sequence and not necessrily of all. 
As another major contribution, we generalized all
sufficient data-independent termination conditions that were known so far.
Our results on chase termination directly carry over to applications that
rely on the chase, such
as~\cite{mms1979,jk1982,bv1984,h2001,l2002,fkmt2006,fkmp2005,dpt2006,ohk2009},
and also to the so-called core-chase presented in~\cite{dnr2008}.
As a sample application, we applied our novel concepts in the
context of~\cite{cgk2008}, showing that they can be used to identify
a larger set of TGDs for which the methods in that paper apply.

There are some interesting open questions left. First, it is unknown
if the membership test for $T[k]$, which has been shown to be in $\textsc{coNP}$,
is also $\mbox{coNP}$-complete. Second, it is left open if $\bigcup_{k \geq 2} T[k]$
is still decidable. Finally, it is an interesting question if the
positive results on core computation in data exchange settings from~\cite{gn2008}
extend to the $T$-hierarchy.

\bibliographystyle{abbrv}
\bibliography{main}  

%\cleardoublepage
%\appendix
%\input{appendix_proofs}

\end{document}